\documentclass[manuscript]{aastex}
\usepackage{bm}
\usepackage[dvipdfmx]{color}  

\slugcomment{Not to appear in Nonlearned J., 45.}

\shorttitle{Formation of multiple-satellite systems}
\shortauthors{Hyodo, Ohtsuki \& Takeda}

\begin{document}


\title{Formation of Multiple-Satellite Systems From Low-Mass Circumplanetary Particle Disks}


\author{Ryuki Hyodo$^1$, Keiji Ohtsuki$^1$ and Takaaki Takeda$^2$}
\affil{$^1$Department of Earth and Planetary Sciences, Kobe University, Kobe
657-8501, Japan; \email{ryukih@stu.kobe-u.ac.jp, ohtsuki@tiger.kobe-u.ac.jp} 
$^2$VASA Entertainment Co. Ltd.}


\begin{abstract}
Circumplanetary particle disks would be created in the late stage of planetary formation 
either by impacts of planetary bodies or disruption of satellites or passing bodies, 
and satellites can be formed by accretion of disk particles spreading across 
the Roche limit. Previous N-body simulation of lunar accretion focused on the 
formation of single-satellite systems from disks with large disk-to-planet mass 
ratios, while recent models of the formation of multiple-satellite systems from 
disks with smaller mass ratios do not take account of gravitational interaction 
between formed satellites. In the present work, we investigate satellite accretion 
from particle disks with various masses, using N-body simulation. In the case 
of accretion from somewhat less massive disks than the case of lunar accretion, formed 
satellites are not massive enough to clear out the disk, but can become 
massive enough to gravitationally shepherd the disk outer edge and start 
outward migration due to gravitational interaction with the disk. When the radial 
location of the 2:1 mean motion resonance of the satellite reaches outside the 
Roche limit, the second satellite can be formed near the disk outer edge, and 
then the two satellites continue outward migration while being locked in the 
resonance. Co-orbital satellites are found to be occasionally formed on the 
orbit of the first satellite. Our simulations also show that stochastic nature 
involved in gravitational interaction and collision between aggregates in the tidal 
environment can lead to diversity in the final mass and orbital architecture, 
which would be expected in satellite systems of exoplanets.
\end{abstract}



\section{Introduction}
\indent There are a variety of satellite systems in our solar system. The terrestrial
planets have no or a small number (one or two) of satellites, while the giant 
planets have many. It is known that the systems of regular satellites of the giant planets have a common ratio of 
the total satellite mass to the host planet's mass with ${\cal O}(10^{-4})$ \cite[e.g.][]{CW06}. 
On the other hand, the satellite-to-planet mass ratio is as large as 0.012 and 
0.1 for the Earth-Moon system and the Pluto-Charon system, respectively. 
Since processes of satellite formation are thought to be closely related to the 
formation process of the host planets, understanding of satellite formation is 
expected to provide constraints on planet formation.\\
\indent The principal regular satellites of the gas giant planets are thought to be formed
by accretion of solids in gas disks around forming host planets \citep[e.g.][]{CW02,CW06,CW09,Es09}. 
For example, \cite{CW06} showed that the common ratio between the total satellite mass and the 
host planet's mass can be naturally explained by a balance between the supply 
of solids into the disk and the loss of formed satellites by inward migration to the 
planet due to the tidal interaction with the gas disk. \cite{OI12} 
demonstrated that the configuration of the Laplace resonance among the inner 
three Galilean satellites can be explained when the existence of the disk inner 
edge is taken into account \citep[see also][]{Sa10}.\\
\indent On the other hand, the Moon is thought to be formed by accretion from an
impact-generated debris disk \citep{HD75, CW76}. 
N-body simulation of lunar accretion from particulate disks showed 
that generally a single satellite is formed in a few months from initial disks of 
masses $M_{\rm disk,ini}/M_{\oplus} \simeq 0.02 - 0.05$ confined within the planet's 
Roche limit \citep{Ida97, Kokubo00}, although the formation time 
scale becomes as long as $\sim 10^2$ years if a hot fluid disk composed 
of vapor and melt is considered, because in such a case the rate of radial 
spreading of the disk and thus the growth rate of the satellite are controlled by 
the rate of radiative cooling of the disk \citep{SC12}. Charon may 
also have been formed by accretion from an impact-generated debris disk, but 
impact simulations showed that it is more likely that Charon is an impactor that 
survived nearly intact \citep{Canup05}. In the above studies of satellite accretion 
from impact-generated disks, relatively massive initial particulate disks are 
necessary to account for the current total mass and angular momentum of the 
Earth-Moon or the Pluto-Charon system; the required disk mass ratio relative to 
the host planet is at least a few percent for the Moon, and more than 10\% for 
Charon \citep{Ida97, Canup05}.\\
\indent On the other hand, \cite{Cha10} considered satellite accretion from a 
circumplanetary particulate disk with a much smaller disk to host planet mass 
ratio to examine formation of moonlets around Saturn. Such a disk would 
have been formed either by tidal disruption of a passing comet \citep{Dones91}, 
disruption by meteoroid impact of a satellite that formed in the circumplanetary 
gas disk and migrated inward past the Roche limit \citep{Cha09}, or 
tidal stripping of the outer layers of a migrating differentiated satellite in the circumplanetary 
gas disk \citep{Canup10}. When satellites are formed outside the Roche limit 
through radial diffusion of disk particles initially confined within the Roche 
limit, the mass of the formed satellite is larger when the disk surface density is 
larger, while the surface density gradually decreases due to the diffusion. As a 
result, satellites that were formed early in the evolution from a disk with a larger 
surface density are more massive. Also, the formed satellites migrate outward 
due to tidal interaction with the particle disk and the planet, but the tidal torques are 
increasing functions of mass. Consequently, more massive satellites migrate outward more 
rapidly, and these different migration rates lead to orbital crossings and merging. 
\cite{Cha10} showed that the above mechanisms can explain the observed mass-orbital 
radius relationship among inner small moons of Saturn (i.e., their mass 
increases with increasing distance from Saturn), if radial spreading of a particle 
disk initially much more massive than the current rings is considered.\\
\indent Furthermore, \cite{CC12} developed an analytic model for the 
accretion and orbital evolution of satellites from particulate disks with mass 
ratios relative to the host planet much smaller than the case of the lunar 
formation, and applied it to the formation of satellite systems of Saturn and 
other giant planets. Assuming that the disk surface density and the mass flux across the 
Roche limit are kept constant, Crida and Charnoz analytically investigated the 
growth and orbital evolution of formed satellites, and found that the observed 
relationship between the masses and orbital radii of regular satellites of Saturn, 
Uranus, and Neptune can be explained by their model. 
It should be noted that, in order to account for the current 
radial location of Saturn's mid-sized moons including Rhea, 
the model by \cite{CC12} \citep[see also][]{Cha11} 
had to assume significantly strong tidal dissipation inside Saturn
corresponding to the quality factor $Q \sim 10^3$ as compared to
the often assumed value of $\sim 10^4$, although 
there is a recent study that suggests such strong dissipation
\citep{Lai12}.\\
\indent The analytic model by \cite{CC12} offers a unified understanding
of the formation of satellite systems of terrestrial planets and those of giant 
planets. However, gravitational interaction between formed satellites is not 
taken into account in their model, which is expected to be important when 
formed satellites are massive, while previous N-body simulations of satellite
accretion mostly focused on accretion of a single moon from relatively massive
disks. On the other hand, using N-body simulation, \cite{Takeda02} investigated 
satellite accretion from particulate disks of various mass ratios to the host 
planet, and examined a relationship between the initial disk mass and the mass  
of formed satellites. Also, in order to examine the processes of the 
formation of the second satellite from the remaining disk after the formation 
of the first satellite, he performed simulations starting from a particulate disk 
with a satellite seed placed outside the Roche limit. However, direct N-body 
simulations of consecutive formation of multiple satellites is desirable for better 
understanding of the whole accretion processes, including detailed orbital 
evolution of growing satellites and formation of co-orbital satellites during 
accretion.\\
\indent Many exoplanets have been found so far, and the search for satellite systems
of exoplanets is ongoing \citep[e.g.][]{Ki12,Ki13a,Ki13b,Ki14}. Such 
satellites around exoplanets are also important in the context of astrobiology, 
because some of them would be located in the habitable zone. Thus, 
understanding of the formation processes of satellites from various particulate 
disks is important, not only for the understanding of the origin of satellite 
systems in our solar system, but also for those of exoplanets.\\
\indent In the present work, we perform N-body simulations to investigate the process 
of consecutive formation of first and second satellites from particulate disks 
initially confined within the Roche limit. We examine how the masses of formed 
satellites depend on the initial disk mass, and also examine their orbital evolution 
in detail. In Section 2, we describe our numerical methods. Section 
3 presents results for satellite accretion from massive disks similar to the 
case of lunar accretion, for comparison with results shown in later sections. 
Numerical results of the consecutive formation of first and second satellites from 
lighter disks are presented in Section 4. Section 5 discusses dependence on 
various parameters and diversity of satellite systems formed from 
particulate disks. Our results are summarized in Section 6.

\section{The Model and Numerical Methods}
\label{Model}
\subsection{Initial Conditions}
We examine accretion of satellites from circumplanetary particle disks in a 
gas-free environment by performing global N-body simulation. Collision and 
gravitational interaction between particles are taken into account. As the initial 
condition, we consider particle disks confined within the planet's Roche limit 
given by
\begin{equation}
 a_{\rm R} = 2.456 \left(\frac{\rho}{\rho_{\rm c}} \right)^{-1/3} R_{\rm c},
\end{equation}
where $\rho$ is the material density of particles, and $\rho_{\rm c}$ and $R_{\rm c}$ are 
the central planet's density and radius, respectively. Initially, particles are 
distributed in an annulus with $0.4a_{\rm R} \leq r \leq a_{\rm R}$, and their surface density 
follows a power-law distribution $\Sigma \propto r^{\beta}$; we examine cases 
with $\beta = -1$, $-3$, and $-5$. Initial orbital eccentricities and inclinations of 
particles are assumed to have small values that follow a Rayleigh distribution. 
In the following, we use the non-dimensional angular momentum of the disk given by 
\begin{equation}
	j_{\rm disk,ini} = L_{\rm disk,ini} / \sqrt{GM_{\rm c}R_{\rm c}},
\end{equation}
where $L_{\rm disk,ini}$ is the angular momentum of the initial disk, and 
$M_{\rm c}$ is the mass of the central planet. 
In the case of surface density distribution given above, we have $j_{\rm disk,ini} \simeq 0.83$, $0.78$, and 
$0.73$ for $\beta = -1$, $-3$, and $-5$, respectively. We examine cases of various initial disk 
masses, with $M_{\rm disk,ini}/M_{\rm c} = 0.01 - 0.06$. The initial disks are described by $3 
\times 10^4$ to $5 \times 10^4$ particles (Table 1).\\
\indent We set the mass of the central planet to be the Earth's mass and assume that
the densities of the particles and the central planet are $\rho = 3.3$ g cm$^{-3}$ 
and $\rho_{\rm c} = 5.5$ g cm$^{-3}$, respectively, so that the case of massive disks 
corresponds to that of lunar accretion examined by previous works \citep{Ida97, Kokubo00}. 
However, evolution of self-gravitating particle disks 
is governed by their surface density rather than the density or mass of individual 
particles, and the mass of the initial disk relative to the planet ($M_{\rm disk,ini}/
M_{\rm c}$) and the initial angular momentum distribution of the disk are important 
parameters that control the outcome of satellite accretion \citep{Ida97, Kokubo00, TI01}. 
Also, $R_{\rm c}/a_{\rm R} \simeq 0.34$ when 
the above values are assumed for the densities of the central planet and the 
particles, while we have $R_{\rm c}/a_{\rm R} \simeq 0.4 - 0.5$ if the densities of the central 
planet and satellites corresponding to the gas giant planets are assumed. 
However, as we show later, dynamical behavior near the disk outer edge 
is important for satellite accretion from a disk initially confined within the Roche 
limit \citep[see also ][]{Ida97, Kokubo00}. Therefore, our results can be
applied to satellite accretion from particle disks in various cases, including ones 
around giant planets.\\
\indent In our model, we assume a disk of condensed particles as the initial 
condition, but this may not be appropriate for massive disks.
In fact, \cite{SC12} showed that vaporization of
constituent particles would be important in the protolunar disk 
around the Earth when $M_{\rm disk}/M_\oplus \gtrsim 0.003$, because
in this case the amount of energy released during the viscous spreading
of the disk is comparable to the latent heat of vaporization of silicate
and the rate of radiative cooling from the disk surfaces is too small 
compared to the viscous heating rate. By generalizing their estimates
to the case of particle disks around a central planet with mass $M_{\rm c}$, 
we find that the amount of energy released during disk spreading is
comparable to the latent heat of vaporization of silicate when 
$M_{\rm c}/M_\oplus \gtrsim 0.3$. On the other hand, the rate of radiative cooling 
from the disk surfaces becomes smaller than the viscous heating rate 
when $M_{\rm disk}/M_\oplus \gtrsim 0.003 \times (M_{\rm c}/M_\oplus)^{-1/3}$.
If we compare the latter condition with the range of disk-to-planet mass
ratios examined in the present work ($0.01 \leq M_{\rm disk,ini}/M_{\rm c} \leq 0.06$),
we find that vaporization would be important at least near the midplane
even in the case of the lightest disk ($M_{\rm disk,ini}/M_{\rm c} = 0.01$) when
$M_{\rm c}/M_\oplus \gtrsim 0.03$. \cite{SC12} showed that the time 
scales of disk spreading and satellite accretion become significantly 
longer when the fluid disk inside the Roche limit is considered.
Also, because of the prolonged period of interaction with the fluid 
inner disk, the formed satellite tends to have a larger semi-major axis
compared to the pure N-body simulations. However, other basic 
characteristics of the final outcomes were quite similar to those of
pure N-body simulations. For example, they found that the relationship
between the mass of the formed satellite and the initial disk angular
momentum can be explained by the semi-analytic relationship derived
from pure N-body simulations \citep{Ida97} with the above
revised estimate of the semi-major axis of the formed satellite.
While we acknowledge the importance of vaporization, in the present
work we will focus on disks of solid bodies, as a first step of full 
N-body simulation of the multiple-satellite system formation.

\subsection{Numerical Methods}
\label{Numerical}
The orbits of particles are calculated by numerically integrating the following 
equation of motion,
\begin{equation}
\frac{d\bm{v}_{i}}{dt} = -GM_{\rm c} \frac{\bm{x}_{i}}{|\bm{x}_{i}|^3} 
	- \sum^{N}_{j \ne i} G m_{j} \frac{\bm{x}_{i} - \bm{x}_{j}}{|\bm{x}_{i} - \bm{x}_{j}|^3},
\end{equation}
where $\bm{x}_i$, $\bm{v}_i$, and $m_i$ denote the position relative to the center of the planet, 
the velocity, and the mass of 
particle $i$, respectively, and $G$ is the gravitational constant. We use the 
modified fourth-order Hermite scheme \citep{KM04} with 
shared time steps for the integration. Gravitational forces between particles 
and the planet are calculated using GRAPE-DR, which is a special purpose 
hardware for gravity calculation.\\
\indent Collision between particles is taken into account, assuming that particles are
smooth spheres with normal restitution coefficient $\varepsilon_{\rm n} = 0.1$. Previous 
works show that dynamical evolution is hardly affected by $\varepsilon_{\rm n}$, as long 
as $\varepsilon_{\rm n} < 0.6$ \citep[e.g.,][]{TI01}. For the search of colliding 
pairs, we adopt the octree method implemented in REBOUND \citep{RS12}. 
When a collision is detected, velocity change is basically calculated 
based on the hard-sphere model \citep{DR94}. Particles that collide 
with the planet are removed from the system. When a collision between 
particles takes place outside the Roche limit, colliding particles can become 
gravitationally bound and form aggregates. Whether colliding pairs become 
gravitationally bound or not is judged by the accretion criteria 
that takes account of the tidal effect \citep{Oh93, Kokubo00}; we 
calculate the energy for the relative motion of the colliding pair using their 
positions and velocities, and the pair is regarded as being gravitationally bound 
when their centers are within their mutual Hill radius and the energy is negative 
\citep{Kokubo00}. Gravitational aggregates formed outside the Roche limit 
can be tidally disrupted when they are scattered back inside the Roche limit 
by a forming satellite or another aggregate. In order to correctly describe such 
phenomena, we basically adopt the so-called “rubble-pile model” \citep{Kokubo00} 
for aggregates formed near the Roche limit, and follow the orbits of 
constituent particles.\\
\indent However, if we adopt this method for all the bodies for the entire simulation, a 
large portion of computing time has to be used for resolving collisions among 
constituent particles that form aggregates, and it is difficult to handle long-term 
orbital integration of formed aggregates. On the other hand, as we will show 
below, a satellite formed by particle accretion near the disk edge gradually 
migrates outward due to gravitational interaction with the disk. Once the 
satellite migrates sufficiently far from the Roche limit, it does not need to be 
treated as a rubble-pile, because it will not be scattered back inside the Roche 
limit any more. Therefore, in the present work, for the largest and the second 
largest aggregates in the system, we adopt the following treatment for the 
merger of constituent particles into a single body. The conditions for this 
treatment are: (i) the aggregate (the largest or the second largest) is 
sufficiently massive, and (ii) the radial distance of its center of mass from the 
planet is sufficiently larger than $a_{\rm R}$ so that it is ensured that it does not 
enter the Roche limit any more. When both the above two conditions are 
satisfied, the aggregate is replaced by a single sphere that has the same mass 
and center-of-mass velocity as those of the aggregate and the density 
equal to that of the constituent particles, $\rho$. The critical values of the mass 
and semi-major axis for this treatment are determined empirically; typically, in 
the case of the first satellite, the critical mass corresponds to about 70\% of its 
final mass (the final mass is estimated empirically from previous works on lunar 
accretion as well as our simulations with and without the above treatment; see 
Figure \ref{Ms_Md}), and the critical distance is about $1.15a_{\rm R}$. When the newly formed 
body (satellite) further accretes other particles, the mass of the satellite is 
increased accordingly.\\
\indent Each simulation is continued until the system reaches a quasi-steady state,
although formed satellites continue slow outward migration 
after their formation. In the case of the formation of a single satellite from massive disks,
simulations are typically continued for 500$T_{\rm K}$ ($T_{\rm K}$ is the orbital period at $r = a_{\rm R}$).
On the other hand, in the case of low-mass disks where the evolution
of the system is slow, we continue our simulations for $10^3 T_{\rm K}$.

\section{Formation of Single-Satellite Systems from Massive Disks}
\label{Single_Satellite}
\begin{figure}
\epsscale{.60}
\plotone{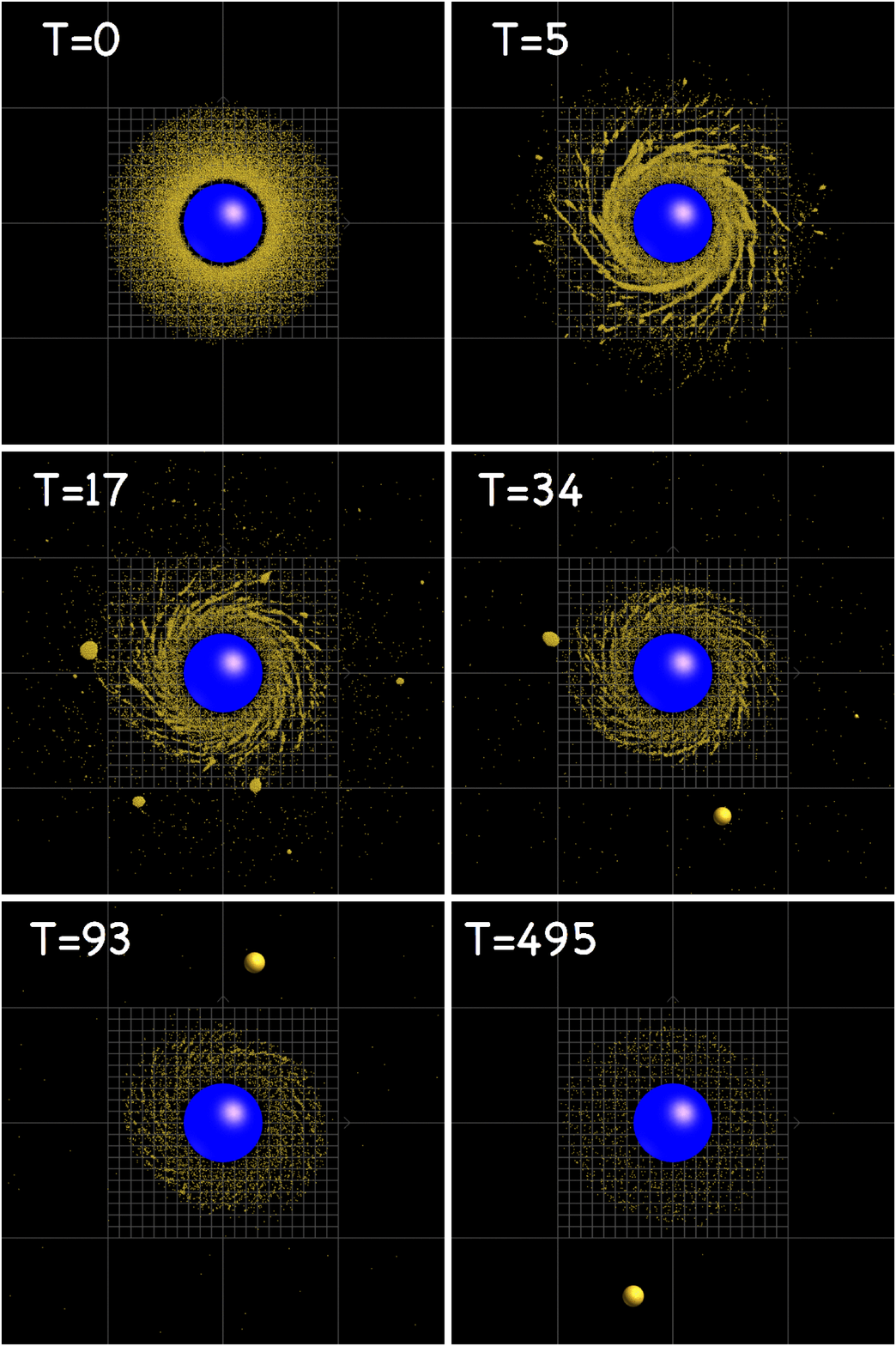}
\caption{Snapshots of the simulation for the case of a single-satellite system 
		(Run-4; $M_{\rm disk,ini}/M_{\rm c} = 0.05$, $j_{\rm disk,ini} = 0.775$). 
		Results are shown looking down onto 
		the circumplanetary particle disk. Initially, the disk is confined in an annulus with $0.4 \leq r/a_{\rm R} \leq 1$, 
		while the radius of the central planet is $R_{\rm c} = 0.34a_{\rm R}$. Numbers in each panel represent the time 
		elapsed since the start of the simulation, in units of the orbital period at the Roche limit.}
\label{Md0.05}
\end{figure}

\begin{figure}
\epsscale{.80}
\plotone{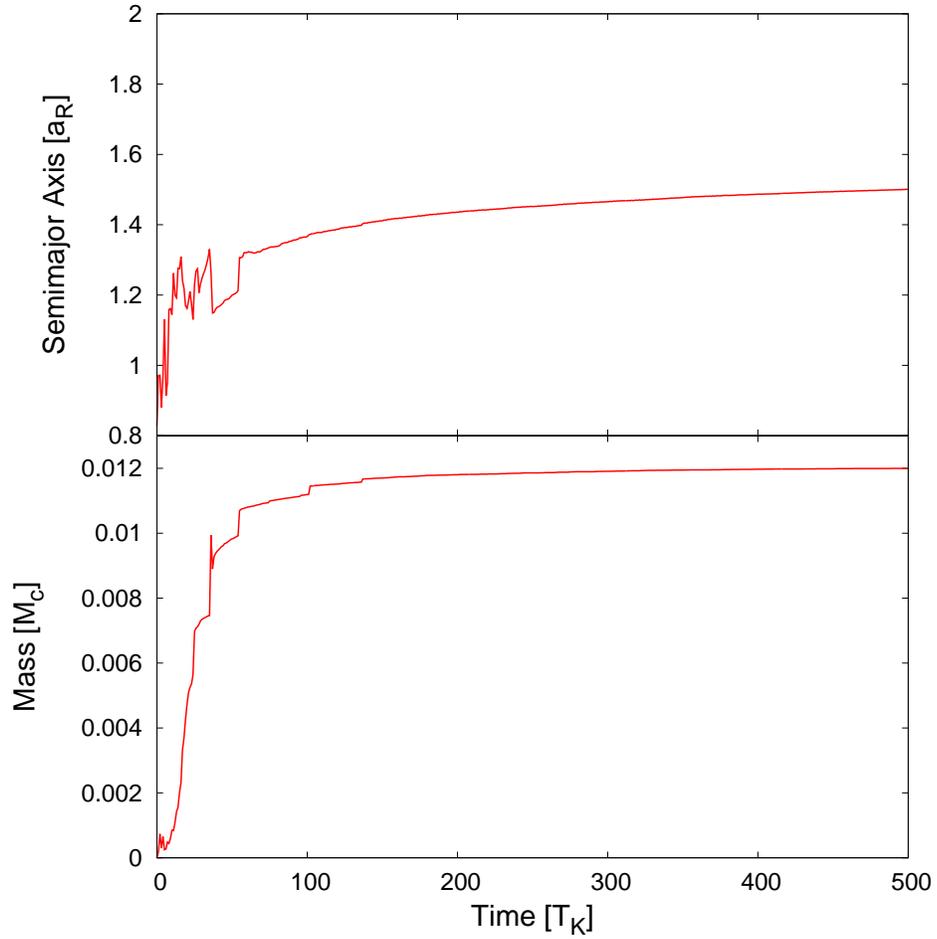}
\caption{Evolution of the semi-major axis (top panel) and the mass (bottom panel) of the 
		satellite in the case of Run-4 ($M_{\rm disk,ini}/M_{\rm c} = 0.05$), where the final outcome is a 
		single-satellite system.}
\label{Md0.05_a_m_evo}
\end{figure}

\begin{figure}
\epsscale{.60}
\plotone{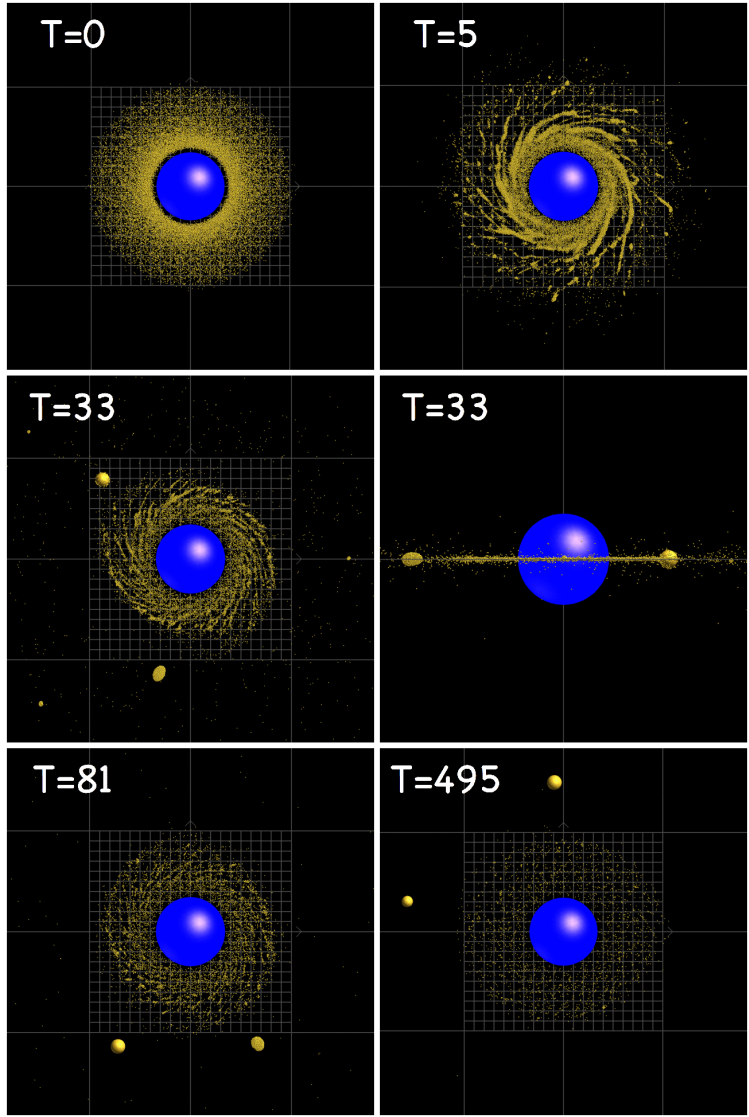}
\caption{Snapshots of the simulation for the case where a system of a large satellite 
		with a small co-orbital one was formed (Run-6; $M_{\rm disk,ini}/M_{\rm c} = 0.04$, $j_{\rm disk,ini} = 0.775$). 
		An edge-on view of the system is also shown for $t=33t_{\rm K}$, where we can see the radially elongated shape of the co-orbital satellite.}
\label{Md0.04}
\end{figure}

\begin{figure}
\epsscale{.80}
\plotone{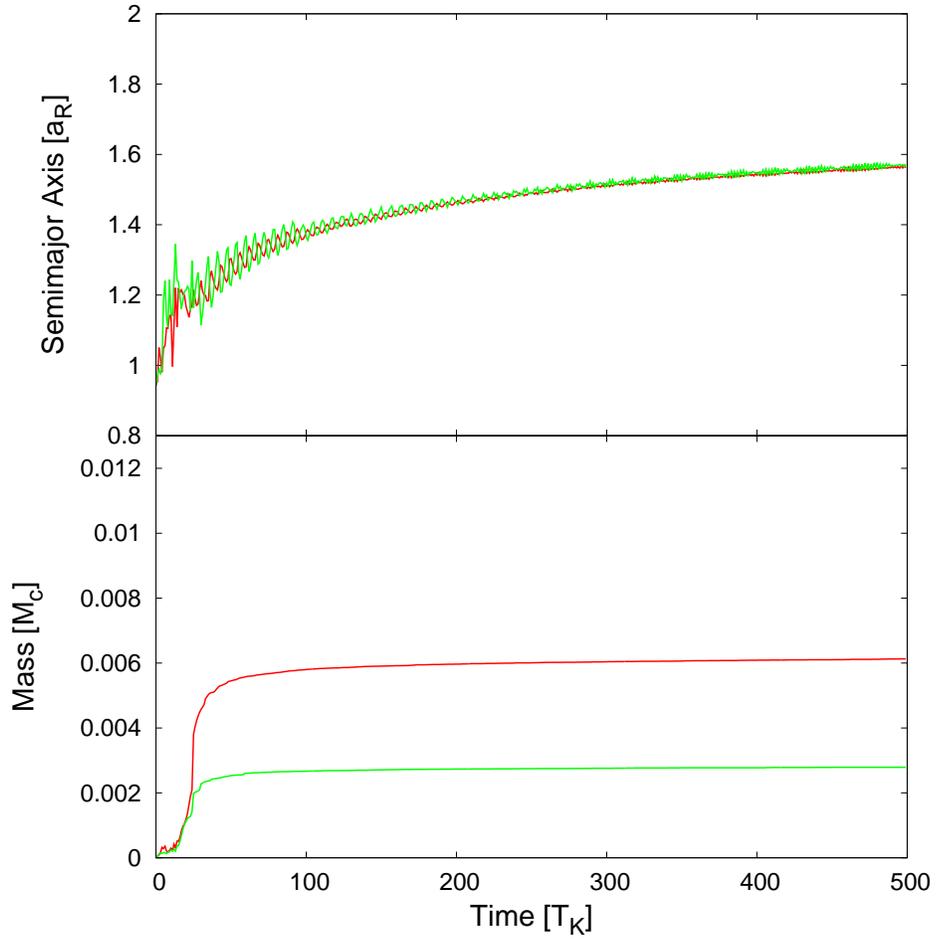}
\caption{Evolution of the semi-major axis (top panel) and the mass (bottom panel) of 
		the satellites in the case of Run-6 ($M_{\rm disk,ini}/M_{\rm c} = 0.04$). The red lines represent 
		those for the primary satellite, and the green lines represent those for the co-orbital satellite.}
\label{Md0.04_a_m_evo}
\end{figure}

\indent In this section, we present results in the case of satellite accretion from 
relatively massive disks with $M_{\rm disk,ini} / M_{\rm c} \simeq 0.05$, for comparison with 
the case of accretion from lighter disks discussed in later sections. Satellite 
accretion from such massive disks corresponds to previous N-body simulations 
of lunar formation \citep{Ida97,Kokubo00}.\\
\indent Figure \ref{Md0.05} shows an example of our simulation that corresponds to such a
massive disk. As we described in Section \ref{Model}, the outer edge of the initial disk is 
located at $r = a_{\rm R}$ (Figure \ref{Md0.05}, $t=0$; the number in each panel represents time in units of the orbital period $T_{\rm K}$  
at the Roche limit). First, spiral structures develop in the disk due to its self-gravity ($t=5T_{\rm K}$). 
Then, angular momentum is transferred through gravitational 
interaction between spiral arms, and those particles transferred outward past 
the Roche limit start forming gravitational aggregates ($t=17T_{\rm K}$). The largest 
aggregate is replaced by a single body when it becomes massive enough 
and is sufficiently far from the Roche limit as a result of outward migration, as 
we described in Section \ref{Numerical} ($t=34T_{\rm K}$). The formed satellite continues outward 
migration ($t=93, 495T_{\rm K}$), while disk particles are scattered by the satellite and 
collide with the planet \citep{Ida97, Kokubo00}. At the end of the 
simulation, a single-satellite system was formed, with the mass of the remaining 
disk being only about 24\% of the satellite mass, or 0.3\% of the planet mass.\\
\indent Figure \ref{Md0.05_a_m_evo} shows the evolution of the mass and semi-major axis of the satellite 
in this case. We can see that the satellite grows rapidly when its orbit is near the Roche limit (i.e., $r/a_{\rm R} \simeq 1-1.2$). 
The time scale of the evolution in the early stage is determined by 
the angular momentum transfer in the disk \citep{Kokubo00,TI01}. 
The viscosity of self-gravitating particle disks can be written as 
\begin{equation}
	\nu = CG^2\Sigma^2/\Omega^3,
\label{viscosity}
\end{equation}
where the coefficient $C$ depends strongly on the distance from the planet \citep{TI01, Dai01}, 
and also weakly on the elastic properties of particles \citep{Ya12}. The time scale of the increase of the satellite mass 
in the early phase of its rapid growth ($\simeq 10^2 T_{\rm K}$) shown in Figure \ref{Md0.05_a_m_evo} is 
roughly consistent with the time scale of viscous evolution of the disk \citep{Kokubo00}.\\
\indent In the case of massive disks, formation of a single satellite is a typical outcome, 
but in some cases a second largest satellite is formed as a co-orbital satellite 
of the largest one. \cite{Kokubo00} report that in about 10\% of their 
simulations, a co-orbital satellite with mass larger than 20\% of the largest one 
was formed. Figure \ref{Md0.04} shows an example of our simulation that produced such
a co-orbital satellite. In this case, the co-orbital satellite is moving in a tadpole 
orbit about the $L_4$ Lagrangian point of the largest satellite, and it stayed in this 
orbit until the end of the simulation ($t=500T_{\rm K}$). Figure \ref{Md0.04_a_m_evo} shows the evolution 
of the masses and the semi-major axes of the two satellites in this case. We 
find that the secondary satellite is captured into the tadpole orbit during the 
phase of rapid growth of the primary satellite, and is locked in such an orbit 
while it migrates outward with the primary satellite. Collision of such a co-orbital 
satellite onto forming satellites may have played an important role in the impact history of satellites \citep{JA11}.\\
\indent In Figure \ref{Md0.04}, an edge-on views is also shown for the panel for $t=33T_{\rm K}$. This panel 
shows a side-view of the secondary satellite before it is replaced by a single 
body, and we can see that it has a radially elongated shape due to the tidal force. 
Such aggregates formed just outside the Roche limit have elongated shapes, 
and are locked in synchronous rotation. However, with increasing distance 
from the planet as a result of outward migration, the tidal effect weakens and 
aggregates tend to have rounder shapes \citep{KS04}. Since 
in our simulation the largest and the second largest aggregates that satisfy 
the conditions described in Section \ref{Numerical} are respectively replaced by a single 
body, we do not follow the change of satellite shapes in the course of outward 
migration.

\section{Formation of Multiple-Satellite Systems}
\label{Multi_formation}
\subsection{Formation of the Second Satellite}
\indent Next, we present results for accretion from lighter particle disks, where the 
second satellite is formed. In our simulations with $j_{\rm disk,ini} = 0.775$, 
the second satellite was formed when $0.015 \leq M_{\rm disk,ini}/M_{\rm c} \leq 0.03$ (Table 1). 
In the marginal case of $M_{\rm disk,ini}/M_{\rm c} = 0.03$, the second satellite was
not formed in one simulation (Run-7), but it was formed in another 
simulation where initial conditions were generated using a different 
set of random numbers (Run-7b). On the other hand, 
in the case with $M_{\rm disk,ini}/M_{\rm c} = 0.01$, the evolution of the system 
was so slow that we had to stop the simulation before the second 
satellite was formed, although its formation is expected if we continue the simulation. 
Figure \ref{Md0.015_1} shows time series of the evolution of the system in the case 
of $M_{\rm disk,ini}/M_{\rm c} = 0.015$. Initial evolution of the disk is similar to the case of 
more massive disks shown in Section \ref{Single_Satellite}. First, spiral structures are formed as a result 
of gravitational instability ($t=4T_{\rm K}$). Since the critical wavelength of the instability 
($\lambda_{\rm c}$) is proportional to the disk surface density, we find that there are 
a larger number of arms in the present case compared to the case shown 
in Figure \ref{Md0.015_1} ($t=5T_{\rm K}$). As a result, the mass of each aggregate formed from particles 
transferred outside the Roche limit, which is roughly proportional to $\Sigma \lambda_{\rm c} ^2$ 
\citep{Kokubo00}, is smaller compared to the case of more 
massive disks (compare Figure \ref{Md0.015_1}, $t=32T_{\rm K}$ with Figure \ref{Md0.05}, $r=17T_{\rm K}$). Also, since the 
surface density of the disk is smaller, the outward mass flux across the Roche 
limit is also smaller. Collisions between these aggregates outside 
the Roche limit produce the first satellite ($t=69T_{\rm K}$), whose mass ($\sim 0.002 \times M_{\rm c}$) 
is smaller than the case of more massive disks. When this first satellite 
satisfies the conditions described in Section \ref{Numerical}, it is replaced by a single sphere ($t=102T_{\rm K}$).\\
\indent After its formation, the first satellite gradually migrates outward via gravitational 
interaction with the disk, as in the case of more massive disks ($t=69, 102T_{\rm K}$). 
On the other hand, particles near the disk outer edge are pushed inward by 
the satellite, and a gap is formed between the satellite and the disk outer edge 
($t=102T_{\rm K}$). The growth of the satellite almost stalls by this stage, but a significant 
amount of mass corresponding to 3.3 times the satellite mass remains in the 
disk. In the case of more massive disks shown in Section \ref{Single_Satellite}, the mass of the first 
satellite was so large that it gravitationally scatters most of disk particles to lead 
to collision with the planet. However, in the present case of the lighter disk, the first 
satellite is too small to scatter disk particles onto the planet. Instead, the 
satellite migrates outward significantly through gravitational interaction with the 
remaining disk, and then the disk outer edge also gradually expands ($t=565T_{\rm K}$).\\
\indent The torque from the satellite exerts on disk particles at resonant locations \citep{GT80}, 
and there are a number of $m:m-1$ inner mean motion resonances with the first satellite between the orbit of the first satellite 
and the disk outer edge. Because of this resonant effect, particles cannot
spread across the Roche limit immediately after the first satellite is formed 
and starts outward migration. However, these resonances vanish beyond 
the radial location of the 2:1 inner mean motion resonance, which is located 
at $r_{2:1} = 0.63a_{\rm s}$ ($r_{2:1}$ is the radial location of the 2:1 inner mean motion 
resonance with the first satellite, and $a_{\rm s}$ is the semi-major axis of the first 
satellite). Therefore, when $r_{2:1}$ becomes larger than $a_{\rm R}$ as a result of outward migration of the 
first satellite, particles near the disk outer edge pile up outside the Roche limit 
and the accretion of the second satellite begins (Figure \ref{Md0.015_1}, $t=690T_{\rm K}$; \cite{GT78, Takeda02} ).
As in the case of the first satellite, 
the second satellite is replaced by a single sphere when its radial location is 
sufficiently far from the disk outer edge (Figure \ref{Md0.015_2}, $t=986T_{\rm K}$). 
Formation of the second satellite at the location of the 2:1 resonance
with the first one was also found in some cases in the
hybrid simulations of \cite{SC12}, but they mostly
focused on the case of the formation of single-satellite systems
because they were primarily interested in the lunar formation.

\subsection{Orbital Evolution of the First and Second Satellites}
\label{longer_term}
\indent Figure \ref{Md0.015_a_m_evo} shows the evolution of the mass and the semi-major axis of the first 
and the second satellites formed in the case of Run-10. The mass growth of the 
first satellite is similar to the case of the formation of a single satellite from more 
massive disks described in Section \ref{Single_Satellite}; it undergoes rapid growth just outside 
the Roche limit, then repels the disk outer edge and begins outward migration. 
In the present case, the time scale of the viscous evolution of 
the particle disk is longer because of the smaller disk surface density, thus the 
growth time scale of the satellite is also longer. On the other hand, the second 
satellite is formed from particles piled up near the 2:1 resonance with the first 
satellite, as mentioned above. Exactly speaking, the satellite seed, which is formed 
near the location of the 2:1 resonance with the first satellite, grows by accreting 
particles spreading from the disk outer edge, while the orbits of many of these particles 
are interior to the resonance location and have specific angular momentum 
smaller than that of the satellite seed. As a result, the specific angular 
momentum of the second satellite can decrease during its rapid growth, and its 
semi-major axis stays near the disk outer edge (Figure \ref{Md0.015_a_m_evo}). When the second satellite 
becomes sufficiently massive to repel the disk outer edge, its growth stalls and 
outward migration begins. Then the second satellite is captured into the 2:1 
mean motion resonance with the first satellite. Since the angular momentum that 
the second satellite receives from the disk is partly transferred to the outer first 
satellite through resonant interaction \citep{Pea86}, the two satellites continue 
outward migration, with their radial locations being locked in the resonance.\\
\indent A similar evolution can be found in other cases of the formation of two satellites 
from disks with different initial masses. Figure \ref{multiple_a_m_evo} shows results for two 
cases with slightly more massive disks than the case shown above, with the 
same initial non-dimensional angular momentum of the disk. Because of the larger 
surface density of the initial disks, the viscous evolution of the disk and satellite 
formation proceeds faster, but the general feature of the evolution is similar 
to the case shown in Figure \ref{Md0.015_a_m_evo}. In the case of $M_{\rm disk,ini}/M_{\rm c} = 0.025$ (Run-8) shown in 
the right panel of Figure \ref{multiple_a_m_evo}, a co-orbital satellite whose mass is 7\% of the first satellite is formed 
on the orbit of the first satellite. This small companion was captured into the 
tadpole orbit when the first satellite was still near the disk, and migrates outward 
together with the first satellite. Then, when the eccentricity of the first satellite 
grew significantly, its orbit became unstable and collided with the first satellite 
at $t = 833 T_{\rm K}$. Out of our four simulations where two satellites 
(excluding co-orbitals) were formed, capture of a co-orbital satellite by the first satellite was found 
in two cases (Runs 3 and 8), which suggests that the formation of co-orbital satellites would be 
more common than the case of satellite formation from more massive disks \citep{Kokubo00}.  
In the other case (Run-3), the mass of the co-orbital satellite was 7\% of the first satellite. \\
\indent The first and the second satellites continue outward migration while their orbits 
are kept locked in the 2:1 mean motion resonance. In the present work, we do 
not examine their subsequent longer-term evolution in detail. If there is sufficient amount 
of mass still available in the disk when the second satellite migrates sufficiently 
far from the disk outer edge, viscous diffusion of the disk is expected to lead to the 
formation of the third and the forth satellites \citep{CC12}. On the 
other hand, when the first two satellites are rather massive and are locked in 
the 2:1 resonance during their outward migration, as in the case examined in 
the present work, orbital eccentricities of the two satellites can grow during their 
migration, which would influence their subsequent dynamical evolution. Figure \ref{multi_e} 
shows the evolution of the eccentricities of the satellites in the cases where two 
satellites were formed. The eccentricity of the first satellite before the formation 
of the second satellite is generally small. However, after the second satellite 
is formed and is locked into the 2:1 mean motion resonance, the eccentricity 
of the second satellite oscillates and takes on rather large values ($0.1 - 0.25$), 
and the eccentricity of the first satellite increases accordingly ($\sim 0.05$). For 
comparison, we performed three-body orbital integration for the two satellites 
and the planet but without disk particles. The initial conditions were taken from 
those at $t = 800 T_{\rm K}$ of our Run-10, and the integration was continued for 
$5000T_{\rm K}$. We confirmed that the eccentricities of the first satellite ($0.02 - 0.04)$ 
and that of the second satellite ($0.1 - 0.13$) did not grow as large as the cases 
shown in Figure \ref{multi_e}, which suggests that the interaction between the satellites 
and the disk as well as their migration play an important role for the eccentricity 
evolution of the satellites.\\
\indent In the present work, sufficiently massive aggregates that migrated far from the 
disk edge are replaced by single spheres. However, if the eccentricity of the 
second satellite becomes so large that its peri-center gets inside the Roche 
limit, the satellite would undergo tidal disruption, which is not taken into account 
in our simulation after the satellites are replaced by single spheres. Figure \ref{Md0_025_no_replacement} 
shows results of an additional simulation we performed to examine such effects\footnote{This case is not listed in Table 1.}. 
The simulation is the same as Run-8 until the formation of the second satellite, 
but we did not replace it with a single sphere in the present case and continued 
treating it as a rubble pile. We find that the mass of the second satellite grows 
after its formation and then decreases significantly; this is because the satellite 
gets inside the Roche limit after its eccentricity grows significantly, and it 
undergoes tidal disruption. After that, the fragments of the first-generation 
second satellite accrete again to form the second satellite of the second 
generation. Such a cycle of tidal disruption and re-accumulation would be 
repeated. Interestingly, a part of the fragments of the disrupted second satellite 
reach the orbit of the first satellite, and contribute to its additional growth. If the 
first satellite migrates sufficiently outward during such a cycle, the orbit of the 
second satellite would move outward and, eventually, its peri-center would avoid 
getting inside the Roche limit. In any case, orbital evolution of multiple-satellite 
systems is important to understand the final outcomes of satellite accretion from 
particle disks.

\section{Dependence on the Mass and Angular Momentum of the Initial Disk}
\subsection{Disk Evolution and Formation of the First Satellite}
\label{Mass_dependence}
Figure \ref{ms_a_evo} shows the evolution of the mass and semi-major axis of the first 
satellite for various initial disk masses. From Figure \ref{ms_a_evo}(a) we confirm that 
more massive disks lead to more rapid evolution and produce more massive first 
satellites. In general, the satellite grows mostly during the early phase of rapid 
growth, and then its growth slows down. Figure \ref{ms_a_evo}(b) shows the plots of 
the evolution of the semi-major axis of the first satellite for the three cases out 
of the six shown in Figure \ref{ms_a_evo}(a). In all these cases, we can see that the satellite 
undergoes rapid growth when its semi-major axis is smaller than about $1.2a_{\rm R}$ 
by accreting particles spreading across the Roche limit. When the satellite 
grows large enough to repel the disk outer edge, its growth almost stalls. At 
this stage, its semi-major axis is about $1.2-1.3a_{\rm R}$, and its outward migration is 
rather smooth afterwards.

\subsection{Mass of the First Satellite}
\label{first_mass}
Figure \ref{Ms_Md} shows the plots of the mass of the first satellites as a function of the 
initial mass of the disk ($j_{\rm disk,ini}=0.775$). We find that the dependence changes at $M_{\rm disk,ini}/M_{\rm c}  
\simeq 0.03$. In the case of massive disks that produce single-satellite systems, 
\cite{Ida97} analytically showed from the conservation of mass and angular 
momentum that the satellite mass is proportional to the mass of the initial disk 
\citep[see also ][]{Kokubo00}. In the derivation of this relationship, it is assumed 
that particles initially in the disk either accrete to form the satellite, collide 
with the planet, or escape from the system through gravitational scattering 
by the satellite, and that the disk does not remain at the final stage. \cite{Ida97} and 
\cite{Kokubo00} confirmed that their numerical results of N-body 
simulations agree well with the analytic relationship.\\
\indent The dashed line in Figure \ref{Ms_Md} is the fit to our numerical results for $M_{\rm disk,ini}/M_{\rm c} > 0.03$, 
assuming that the satellite mass is proportional to the mass of the 
initial disk. We confirm that our results for the case of the formation of single-satellite 
systems agree well with this relationship as in the previous works. On 
the other hand, in the case of lighter disks with $M_{\rm disk,ini}/M_{\rm c} < 0.03$, a significant 
mass still remains in the disk after the formation of the first satellite, and the 
second satellite is formed from the remaining disk. Thus, the assumption of 
the complete depletion of the initial disk at the time of the formation of the first 
satellite that was made in the derivation of the above analytic linear relationship 
between the satellite mass and the initial disk mass is not valid anymore. In fact, 
in the case of such low mass disks, the dependence of the mass 
of the first satellite on the initial disk mass is found to be stronger, and the satellite mass is 
approximately proportional to the square of the initial disk mass. In this case, the 
accretion efficiency of incorporation of disk material into the satellite \citep{Kokubo00} 
decreases with decreasing disk mass.\\
\indent On the other hand, satellite accretion from particle disks with much lower initial 
mass was studied by \cite{CC12}, as we mentioned before. They 
investigated satellite accretion in the limit where the mass of the satellites accreted 
from the disk is negligible compared to the disk mass, and assumed that the 
disk surface density and the mass flow rate across the Roche limit are constant 
during satellite accretion. In this case, if we define the time scale of the viscous 
spreading of the disk with viscosity $\nu$ as $T_{\rm \nu} \equiv a_{\rm R}^2/\nu$, the rate of 
outward mass flow across the disk outer edge can be written as $F = M_{\rm disk}/
T_{\rm \nu}$. Using the expression of the viscosity for the self-gravitating collisional 
disks (Equation \ref{viscosity}) and $M_{\rm disk} = \pi a_{\rm R}^2\Sigma$, we then have 
\begin{equation}
	F=\pi C G^2 \Sigma^3/\Omega^3.
\end{equation}
In the model of \cite{CC12}, satellites (or satellite seeds) formed 
near the disk outer edge grow by directly accreting particles spreading from the 
disk outer edge when the satellites are still near the edge ("continuous regime"), 
or by capturing moonlets formed by accretion of such spreading particles when 
the satellites somewhat migrate outward ("discrete regime"). In both cases, the 
growth rate of the mass of the satellites ($M_{\rm s}$) is determined by the mass flow rate $F$, and we have 
\begin{equation}
	M_{\rm s} \propto F \propto \Sigma^3 \propto M_{\rm disk}^3.
\end{equation}
\cite{CC12} showed that the mass of the formed satellite at the end of the discrete regime is given by 
$M_{\rm s}/M_{\rm c} \simeq 2200 (M_{\rm disk}/M_{\rm c})^3$. 
These satellites continue outward migration due to torques from the particle 
disk and the planet, and grow further by mutual collisions in the course of the 
migration ("pyramidal regime"). \\
\indent In Figure \ref{differents}, we compile the fitting results to our simulations as well as those 
of previous works for satellite accretion from massive disks \citep{Ida97, Kokubo00} 
and from light disks \citep{CC12}. As we 
mentioned above, in the case of the formation of single-satellite systems from 
massive disks, the disk is cleared out quickly due to gravitational scattering 
by the massive first satellite. In this case, the mass of the first satellite is 
proportional to the initial disk mass \citep{Ida97, Kokubo00}. On 
the other hand, the assumption of a constant disk surface density seems to be 
reasonable when the mass of the formed satellites is much smaller than the 
disk mass. In this case, the mass of the satellites is proportional to the cube of 
the disk mass, as we have shown above. The case of the formation of the first 
and the second satellites in our simulation is intermediate between the above 
two cases. In this case, the mass of the first satellite is not large enough to clear 
out the remaining disk, while it is too massive to neglect its influence on the 
remaining disk; the mass of the disk significantly decreases as a result of the 
formation of the first satellite, and the disk outer edge is shepherded by the first 
satellite after its formation. As a result, the dependence of the mass of the first 
satellite on the initial disk mass becomes also intermediate between the above 
two cases.

\subsection{Formation of the Second Satellite}
\label{Ld}
In the case of the disks with non-dimensional angular momentum $j_{\rm disk,ini} = 0.775$ shown 
above, the second satellite was formed when $0.015 \leq M_{\rm disk,ini}/M_{\rm c} \leq 
0.03$, while single-satellite systems are formed from more massive disks (Section \ref{Multi_formation}). 
However, evolution of particle disks depends not only on the mass but also on 
the angular momentum distribution of the initial disk \citep{Ida97, Kokubo00}. 
Figure \ref{ms1_or_w_ms2} shows 
results of a series of simulations where the mass and angular momentum of 
the initial disk are varied. The filled circles show the cases where the second 
satellite was formed, while the open circles represent the cases where single-satellite 
systems are formed. The open triangle indicates the marginal case where whether the 
second satellite is formed or not depends on the choice of random numbers for generating 
initial conditions. From Figure \ref{ms1_or_w_ms2}, we find a 
tendency that the critical mass of the initial disk that produces multiple satellites 
decreases with increasing angular momentum, indicating that higher angular 
momentum of the initial disk facilitates the formation of single-satellite systems. 
Figure \ref{Ld_dependence} shows the dependence of the mass 
of the first satellite on the disk angular momentum. 
When the disk angular momentum is larger and more mass is located in the 
outer part of the disk, particles in the disk can be transported outward across 
the Roche limit more easily, thus the mass of the first satellite tends to be larger. 
Such a massive first satellite easily clears out the remaining disk, 
and the second satellite cannot be formed. On the other hand, 
in the case of compact disks with smaller angular momentum, the 
outward mass flux across the Roche limit is rather small and the mass of the 
first satellite tends to be also small, which facilitates the formation of the second 
satellite.

\subsection{Mass of the Second Satellite}
Figure \ref{Ms1Ms2} shows the mass of the second satellites together with that of the first 
satellites as a function of the initial disk mass. As we mentioned in Section \ref{first_mass}, 
the mass of satellites produced from lighter disks tends to be smaller. At the 
time of the accretion of the second satellite, the mass and the surface 
density of the disk are smaller compared to the initial disk due to the formation 
of the first satellite. Therefore, the mass of the second satellite tends to be 
smaller than that of the first satellite.\\
\indent On the other hand, as we mentioned above, the efficiency of incorporation of 
disk material into the first satellite decreases with decreasing disk mass in the 
case of the formation of multiple satellites. When the mass of the first satellite 
is much smaller than the initial disk mass, the mass of the remaining disk at the 
time of the formation of the second satellite is still similar to that of the initial disk. 
As a result, the mass of the second satellite formed from such a remaining disk 
becomes similar to that of the first satellite. From Figure \ref{Ms1Ms2}, we can confirm 
that the difference between the masses of the first and the second satellites 
becomes smaller with decreasing mass of the initial disk. This is also consistent 
with \cite{CC12}, who considered the case with much lighter disks. 
In this case, the mass of the formed satellite at the end of the discrete regime is 
given by $M_{\rm s}/M_{\rm c} \simeq 2200(M_{\rm disk}/M_{\rm c})^3$ as we mentioned above, 
which is constant as the disk mass is assumed to be constant.

\subsection{On the Diversity in Final Outcomes}
\label{diverse_result}
As we have shown above, the typical outcome of satellite accretion from 
massive particle disks is single-satellite systems, while multiple-satellite 
systems are produced from lighter disks. Also, when multiple satellites are 
formed, outer satellites tend to be more massive, because of a larger surface 
density of the disk at the time of accretion and/or as a result of outward 
migration and merger of formed satellites. In both cases of single- and multiple-satellite 
systems, co-orbital satellites can be formed occasionally.\\
\indent On the other hand, dynamical evolution of self-gravitating particle disks 
inevitably involves stochastic nature arising from gravitational scattering 
between aggregates as well as their collisional disruption and subsequent re-accumulation, 
which can result in diversity in outcomes of disk evolution. As an  
example, Figures \ref{Md0.03_diversity} and \ref{diverse} show results of Run-7b; in this case, the parameter 
values of disk mass and angular momentum are the same as those assumed 
in Run-7 ($M_{\rm disk,ini}/M_{\rm c} = 0.03$, $j_{\rm disk} = 0.775$), but a different set of random 
numbers are used to generate initial positions and velocities of particles. In the 
case of Run-7, only one satellite was formed. On the other hand, two satellites 
are formed in Run-7b; unlike the above-mentioned typical cases, the inner 
satellite is more massive than the outer one. Also, the two satellites are found to 
be locked in the 1:2 mean motion resonance.\\
\indent This system with an inner larger satellite and an outer smaller one was formed 
in the following way. First, radial spreading of particles from the disk leads to 
the formation of a number of satellite seeds just outside of the Roche limit, and 
they grow through mutual collision and accretion. However, accretion efficiency 
between bodies in the strong tidal field is not 100\% \citep{Oh93,CE95,Oh13}, 
and collision between gravitational 
aggregates can lead to complete or partial disruption \citep{Kar07,Hyo14}. 
In the case of $M_{\rm disk,ini}/M_{\rm c} = 0.03$ shown in Figures \ref{Md0.03_diversity} and 
\ref{diverse}, the mass of satellite seeds is significantly large. When such aggregates are 
disrupted by collision, they are often elongated due to the tidal effect and 
ejected particles are spread in a rather wide range of radial locations. Also, 
gravitational interaction between aggregates as well as between aggregates 
and dispersed particles leads to significant changes in semi-major axes and 
eccentricities of aggregates. In the case shown in Figure \ref{Md0.03_diversity}, a relatively 
massive aggregate (which eventually becomes the outer smaller satellite in the 
final state) undergoes collision with another aggregate, which leads to partial 
disruption of the colliding bodies (Figure \ref{Md0.03_diversity}, $t=29T_{\rm K}$). The mass of the aggregate shows abrupt 
increase at the time of the collision, because the temporarily combined object is 
regarded as a single body (the green line at $t\simeq28T_{\rm K}$ in Figure \ref{diverse}a). The largest remnant body produced by this collision 
and disruption has a rather large semi-major axis, and its eccentricity is also 
increased (the green line in Figures \ref{diverse}b, c). Then, at t $\simeq 33T_{\rm K}$ (Figure \ref{Md0.03_diversity}), another 
aggregate experiences scattering and partial disruption that leads to outward 
displacement, when the aggregate loses significant mass (the red line in 
Figures \ref{diverse}a, b). Then, this aggregate undergoes a close encounter with the 
aggregate that was scattered outward before, and the former is scattered 
inward to the vicinity of the outer edge of the disk ($t \simeq 43T_{\rm K}$; the red line 
in Figure \ref{diverse}b). Afterwards, this satellite seed grows by accreting particles 
spreading from the disk outer edge, and eventually becomes the largest satellite (Figure 
\ref{Md0.03_diversity}, $t = 59, 146T_{\rm K}$; Figure \ref{diverse}a). During the growth of this largest satellite, 
particles located in the region exterior to the satellite's orbit are scattered by the 
satellite to outer orbits, and then captured by the outer smaller satellite. Thus, 
the outer satellite grows significantly during this phase. When the mass of the 
inner satellite becomes large enough to shepherd the outer edge of the 
disk, the satellite begins outward migration. The radial distance between the 
inner larger satellite and the outer smaller one gradually shrinks and, 
eventually, they get captured into the 1:2 mean motion resonance \citep[see also][]{SC12}. 
Then they continue outward migration while being locked in the resonance. The 
eccentricity of the outer smaller satellite becomes large ($\sim 0.2$) after it is 
captured into the resonance (the green line in Figure \ref{diverse}c). The eccentricity of the inner satellite is rather 
small initially, but it grows significantly in the course of the outward migration 
due to the resonant effect (the red line in Figure \ref{diverse}c).\\
\indent In the above evolution, gravitational scattering between aggregates and their 
imperfect accretion at collision in the tidal environment plays an essential role 
in producing the final outcome that is different from the typical case. When the disk is not too light, as in 
the case shown above, those aggregates formed near the disk outer edge have 
significant mass, and gravitational scattering between them and their disruption 
plays an important role in delivering significant mass to the outer region, which 
facilities formation of satellites with large orbits. Thus, stochastic nature of 
gravitational scattering and collisional disruption in the tidal field can lead to
significant diversity in the final outcome of satellite accretion.

\section{Summary}
In the present work, using N-body simulations, we have investigated formation 
of satellites from particulate disks initially confined within the Roche limit. While 
the formation of single-satellite systems from massive disks was examined 
by previous N-body simulation \citep{Ida97,Kokubo00}, we have 
studied processes of the formation of multiple-satellite systems from low-mass 
disks in detail. When the mass of the particle disk is smaller, a larger number of 
particles have to be used in the simulation to resolve spiral structures created 
due to collective effects among particles, which plays an essential role in the 
disk's dynamical evolution. Also, simulations for a longer period of time are 
required, because the time scale of the disk evolution becomes longer with 
decreasing surface density of the disk. In order to solve these problems, we 
have adopted a new approach; we replace the largest and the second largest 
aggregates by a single sphere, respectively, when they become sufficiently 
massive and migrate outward sufficiently far from the disk outer edge. This 
approach allowed us to perform simulations of consecutive formation of the 
first and the second satellites from particle disks. Formation of multiple-satellite 
systems has been recently examined based on analytic and numerical models 
\citep{Cha10,CC12}, but gravitational interaction between satellites was not taken into account in these studies.\\
\indent We found that single-satellite systems are formed from massive disks, while 
multiple-satellite systems are formed from lighter disks. In the case that the non-dimensional 
disk angular momentum $j_{\rm disk,ini} = 0.775$, multiple-satellite 
systems were formed when $M_{\rm disk,ini}/M_{\rm c} \leq 0.03$, and we found that multiple-satellite 
systems are more likely to be formed from disks with smaller angular 
momentum. In the case of the formation of single-satellite systems from 
massive disks, previous studies showed that the mass of the formed satellite is 
proportional to the initial mass of the disk \citep{Ida97,Kokubo00}. 
On the other hand, the recent analytic study examined satellite accretion in the 
limit where the mass of the satellites accreted from the disk is negligible compared 
to the disk mass, assuming that the disk surface density and the mass flow rate 
across the Roche limit are constant during satellite accretion \citep{CC12}. 
In this case, it has been shown that the mass of the satellites is 
proportional to the cube of the disk mass. Our simulations show that the mass 
of the first satellite in the case of the formation of two-satellite systems is 
approximately proportional to the square of the disk mass for 
$0.01 \leq M_{\rm disk,ini}/M_{\rm c} \leq 0.03$, which is intermediate between the above two cases. We 
also found that the second satellite is formed near the location of the 2:1 mean 
motion resonance with the first satellite. After the formation of the second 
satellite, the two satellites continue outward migration while being locked in 
the resonance \citep[see also][]{SC12}. When the eccentricity of the inner satellite grows significantly as 
a result of this resonant effect, its pericenter would get inside the Roche limit 
and the satellite would undergo tidal disruption, while it would avoid disruption if 
it migrates outward sufficiently far from the Roche limit before the eccentricity 
grow significantly. More detailed studies are required for the orbital evolution of 
multiple-satellite systems formed by accretion from particle disks. On the other 
hand, co-orbital satellites are occasionally formed on the orbits of the first 
satellite \citep{Kokubo00}. Although we did not find formation of co-orbital 
satellites on the orbit of the second satellite, this is likely due to the resolution of 
our simulations. Our results suggest that the observed co-orbital satellites in the 
Saturnian satellite system may have formed during the accretion of the primary 
satellites. Also, collision of co-orbital satellites onto forming satellites may have 
played an important role in the impact history of satellites \citep{JA11}.\\
\indent In most cases of our simulations that produced two-satellite systems, the first 
satellite on the outer orbit was more massive than the second one that  
was formed later on the inner orbit, and the two satellites are locked in the 
2:1 mean motion resonance, as mentioned above. However, our simulation 
also showed that accretion from particle disks can produce satellite systems 
significantly different from such a typical outcome, owing to the stochastic 
nature involved in gravitational interaction and collision between aggregates in 
the tidal environment (Section \ref{diverse_result}). Such effects may have played an important role 
in producing characteristics of the mass and orbital architecture of the satellite 
systems in our Solar System. They would also be important in the formation of 
satellite systems of exoplanets, thus more detailed studies are desirable.

\acknowledgments
We are grateful to Shigeru Ida for discussion and support. We also thank Sebastien Charnoz and Aurelien Crida for discussion, 
and Hiroshi Daisaka for valuable assistance and advice in using GRAPE-DR systems. 
This work was supported by MEXT/JSPS KAKENHI. 
Part of numerical simulations were performed using the GRAPE system at the Center for Computational Astrophysics of the  
National Astronomical Observatory of Japan. Visualization of simulation results was performed in part using Zindaiji 3. 
One of the authors (R. H.) would like to dedicate this work with gratitude to Kazuki Sumi.



\clearpage

\clearpage

\clearpage

\clearpage

\clearpage
\begin{figure}
\epsscale{.90}
\plotone{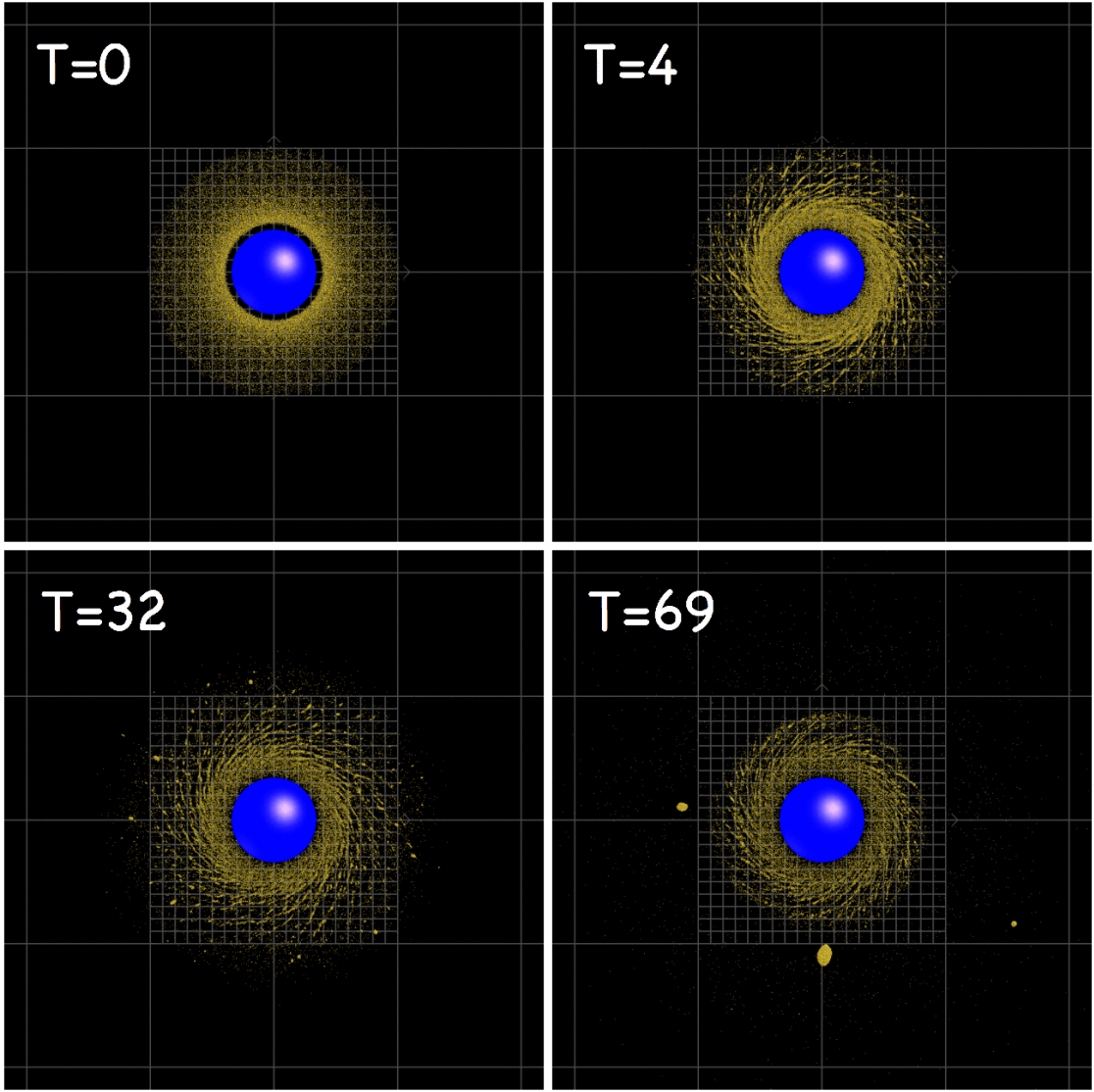}
\caption{Same as Figure \ref{Md0.05}, but the case of the consecutive formation of two satellites is 
		shown (Run-10; $M_{\rm disk,ini}/M_{\rm c} = 0.015$, $j_{\rm disk,ini} = 0.775$).}
\label{Md0.015_1}
\end{figure}

\clearpage
\setcounter{figure}{4}
\begin{figure}
\epsscale{.90}
\plotone{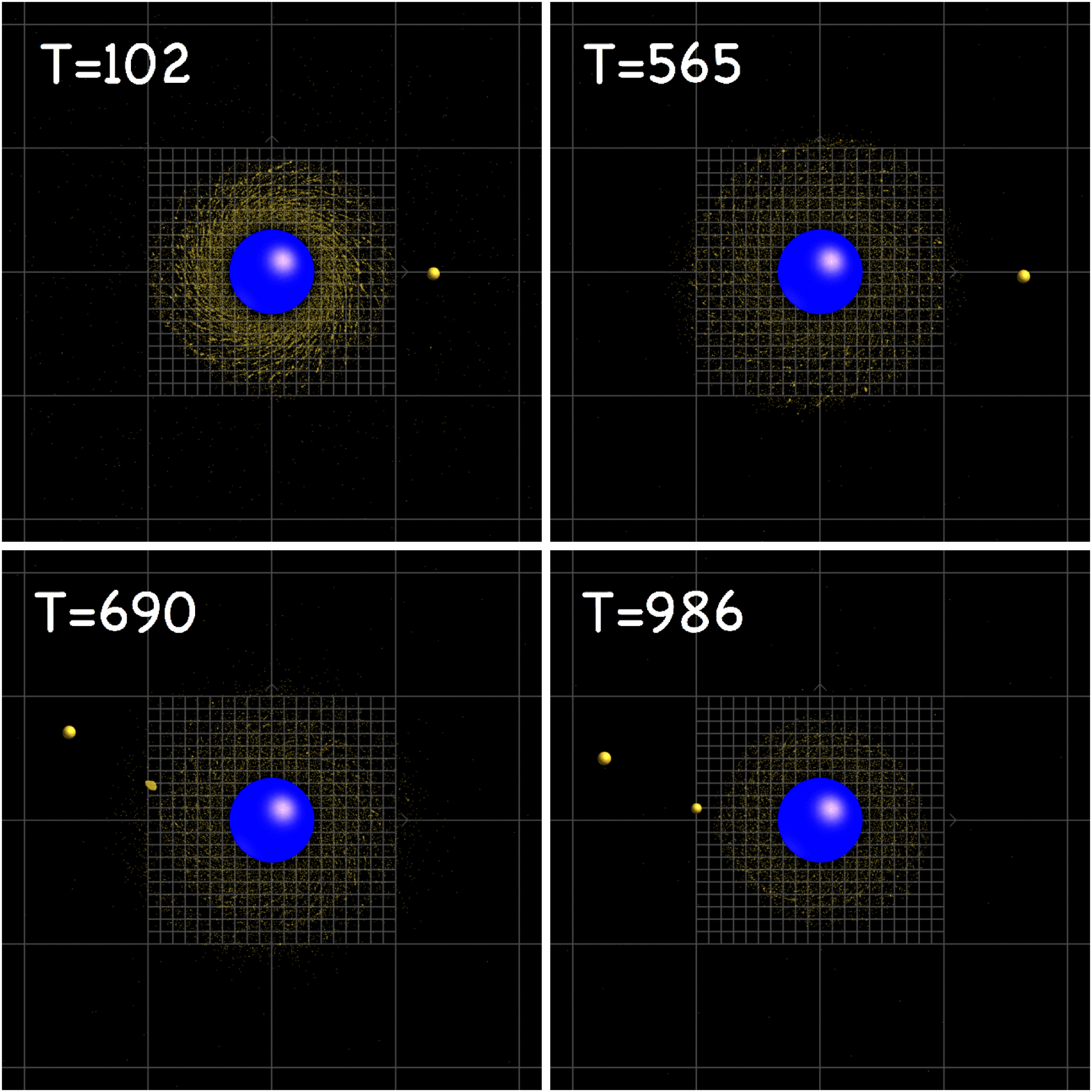}
\caption{{\it Continued.}}
\label{Md0.015_2}
\end{figure}

\clearpage
\begin{figure}
\epsscale{.80}
\plotone{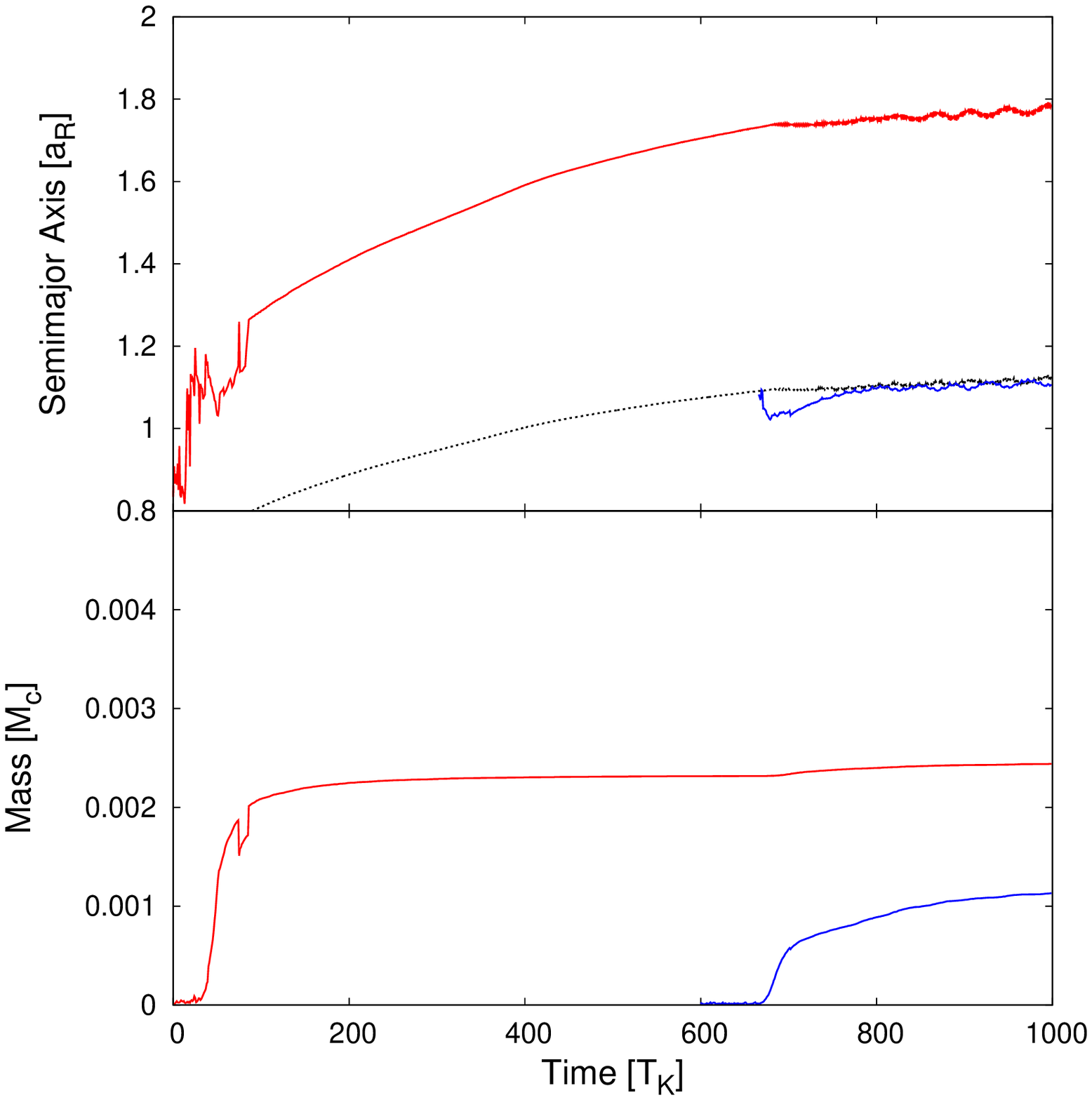}
\caption{Evolution of the semi-major axis (top panel) and the mass (bottom panel) of the 
		satellites in the case of Run-10 ($M_{\rm disk,ini}/M_{\rm c} = 0.015$). The red lines represent 
		the first satellite, and the blue lines represent the second one. The black dotted 
		line in the top panel shows the radial location of the 2:1 mean motion resonance 
		with the outer first satellite.}
\label{Md0.015_a_m_evo}
\end{figure}

\clearpage
\begin{figure}
\epsscale{1.00}
\plotone{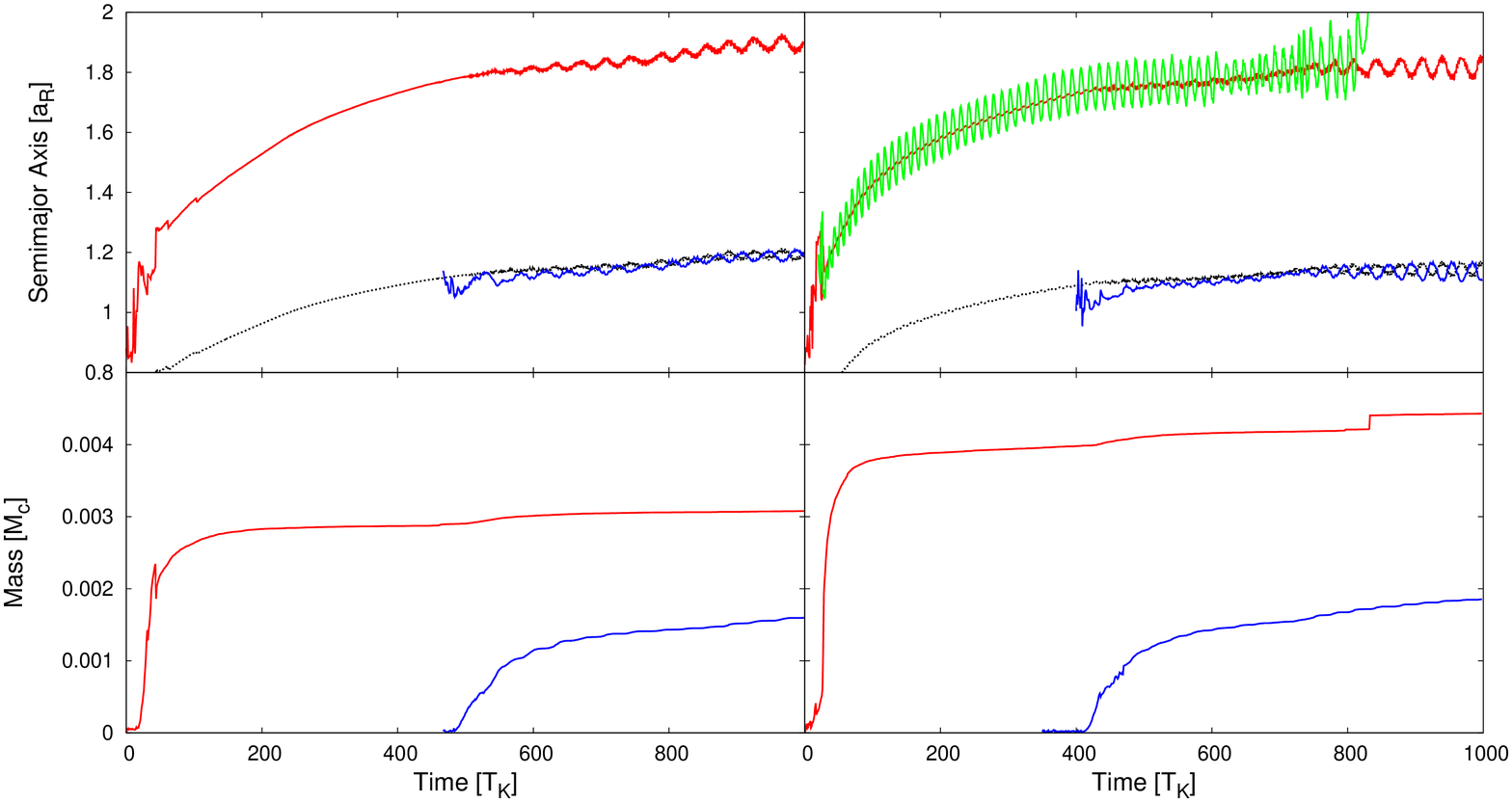}
\caption{Same as Figure \ref{Md0.015_a_m_evo}, but those for Run-9 ($M_{\rm disk,ini}/M_{\rm c} = 0.02$, left panel) and 
		Run-8 ($M_{\rm disk,ini}/M_{\rm c} = 0.025$, right panel) are shown. In the case of Run-8, a 
		co-orbital satellite was temporarily formed (green line), but it eventually collides 
		and merges with the first satellite at $t = 833T_{\rm K}$.}
\label{multiple_a_m_evo}
\end{figure}

\clearpage
\begin{figure}
\epsscale{.50}
\plotone{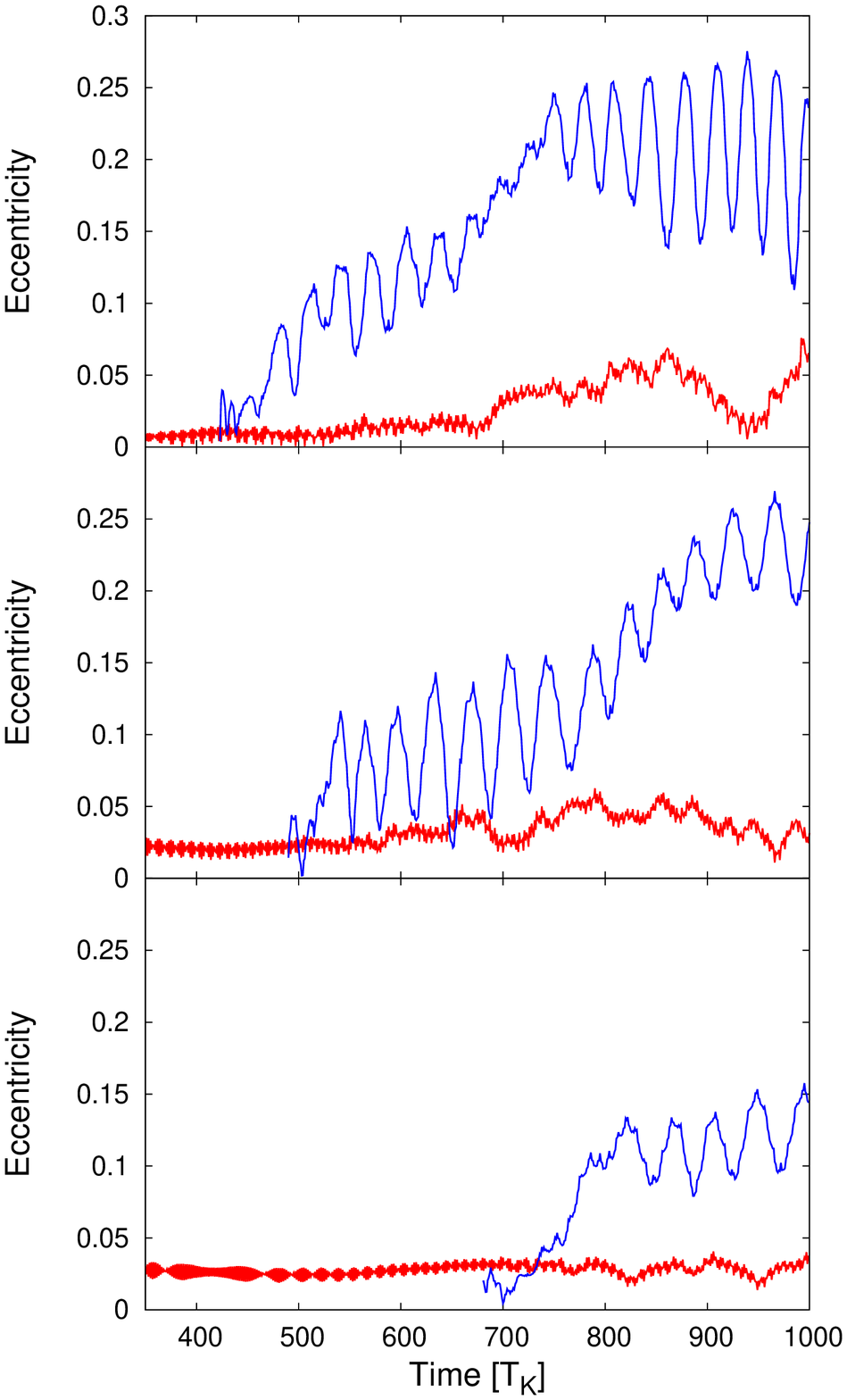}
\caption{Evolution of the orbital eccentricity of satellites for Run-8 ($M_{\rm disk,ini}/M_{\rm c} = 0.025$, 
		top panel), Run-9 (0.02, middle panel), and Run-10 (0.015, bottom panel). The 
		red lines represent the first satellite, and the blue lines represent the second satellite.}
\label{multi_e}
\end{figure}

\clearpage
\begin{figure}
\epsscale{.80}
\plotone{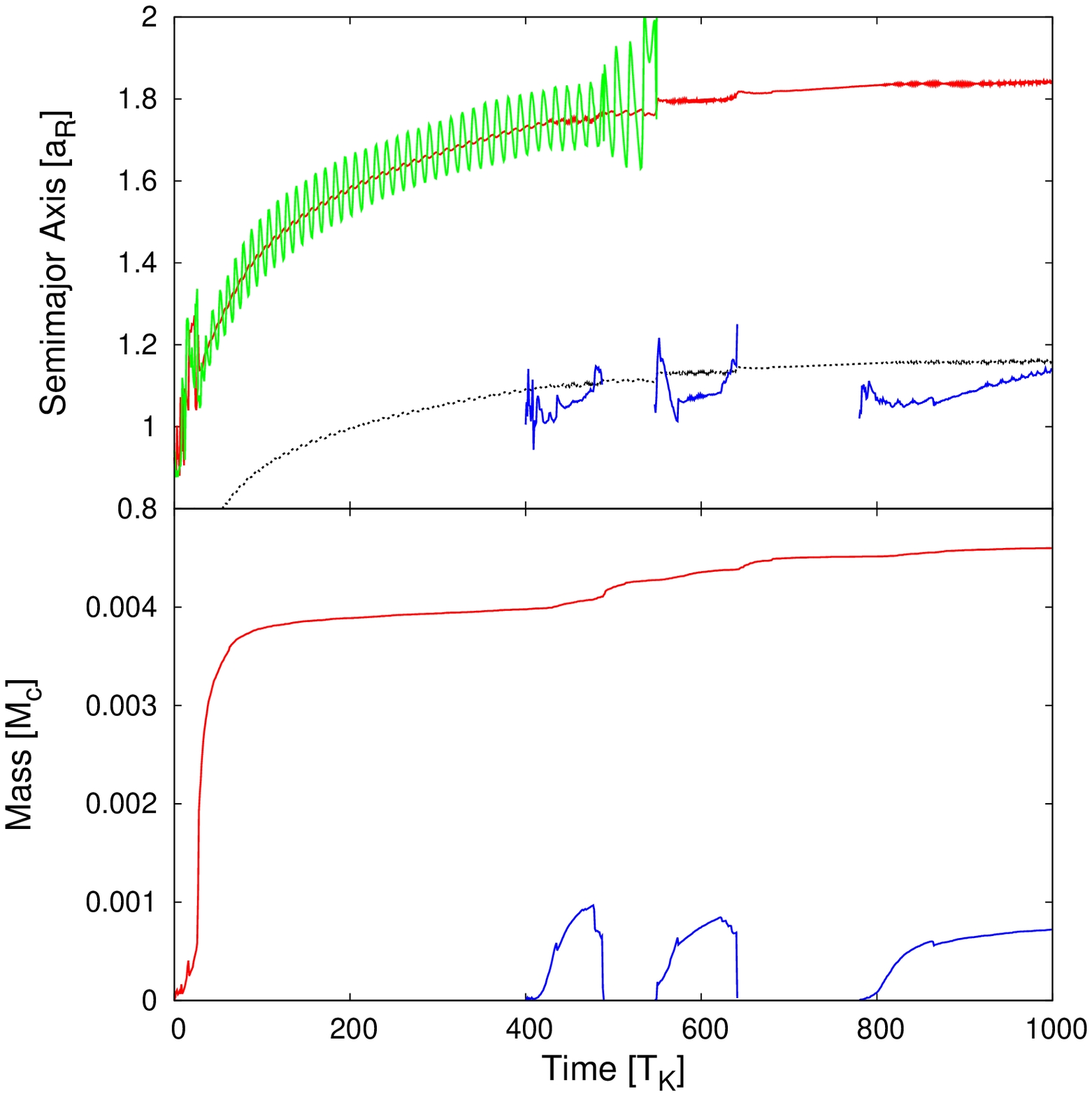}
\caption{Evolution of the semi-major axis (top panel) and the mass (bottom panel) of 
		the satellites in the case of the additional simulation, where the second satellite 
		is treated as a rubble-pile object, without being replaced by a single sphere. 
		Note that the simulation is the same as Run-8 shown in the right panel of 
		Figure \ref{multiple_a_m_evo} until the formation of the second satellite. The meanings of the lines 
		are the same as the right panel of Figure \ref{multiple_a_m_evo}. In the present case, the co-orbital 
		companion of the first satellite is scattered inside the Roche limit at $t \simeq 550T_{\rm K}$.}
\label{Md0_025_no_replacement}
\end{figure}

\clearpage
\begin{figure}
\epsscale{.70}
\plotone{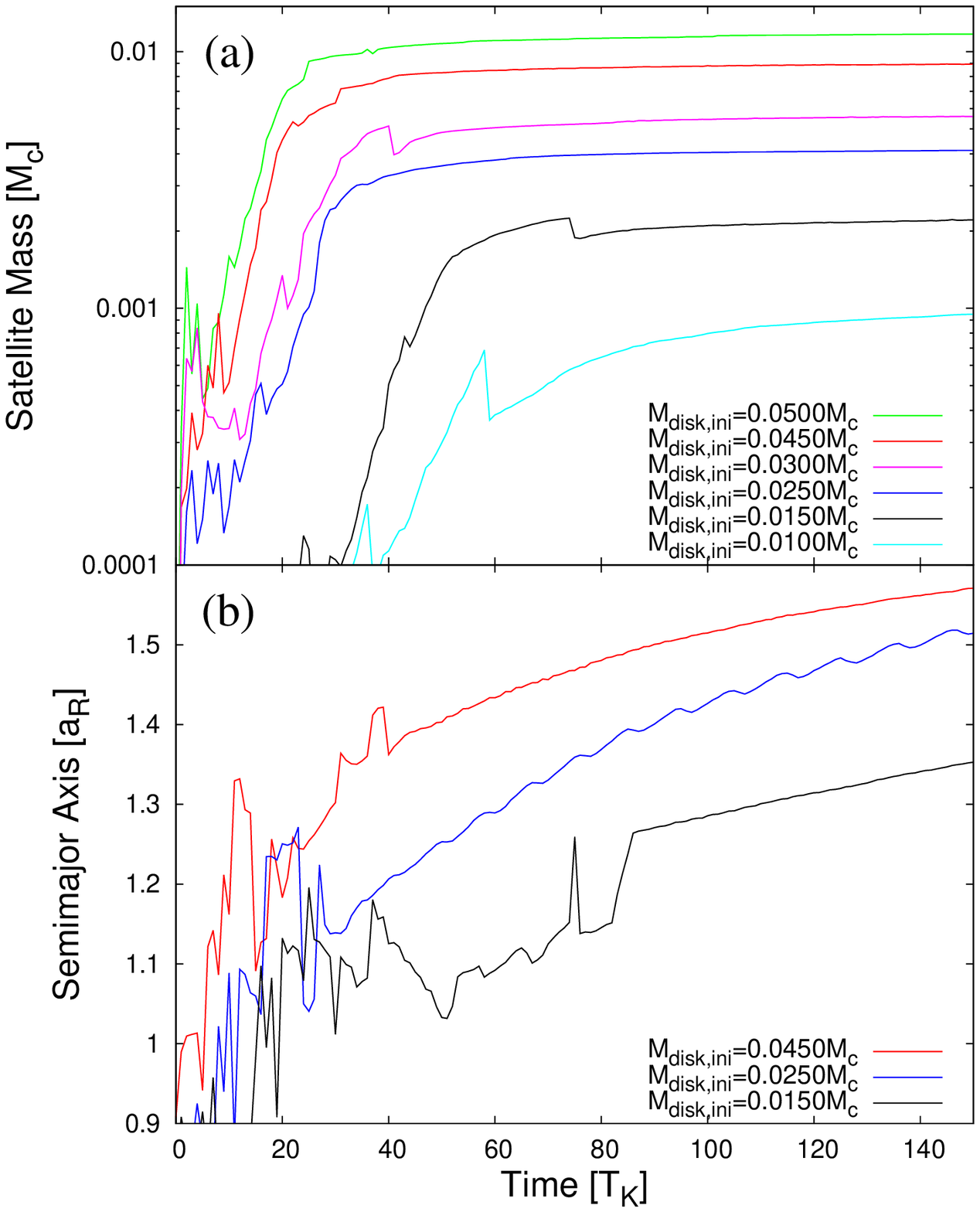}
\caption{Evolution of the mass (top panel) and the semi-major axis (bottom panel) of the 
		largest satellite in cases with various disk masses ($j_{\rm disk,ini} = 0.775$). In the top 
		panel, the mass of the co-orbital companion on the orbit of the largest satellite 
		as well as that of a small moonlet formed on an orbit exterior to the largest 
		satellite's orbit is added to that of the largest satellite to facilitate comparison.}
\label{ms_a_evo}
\end{figure}

\clearpage
\begin{figure}
\epsscale{.80}
\plotone{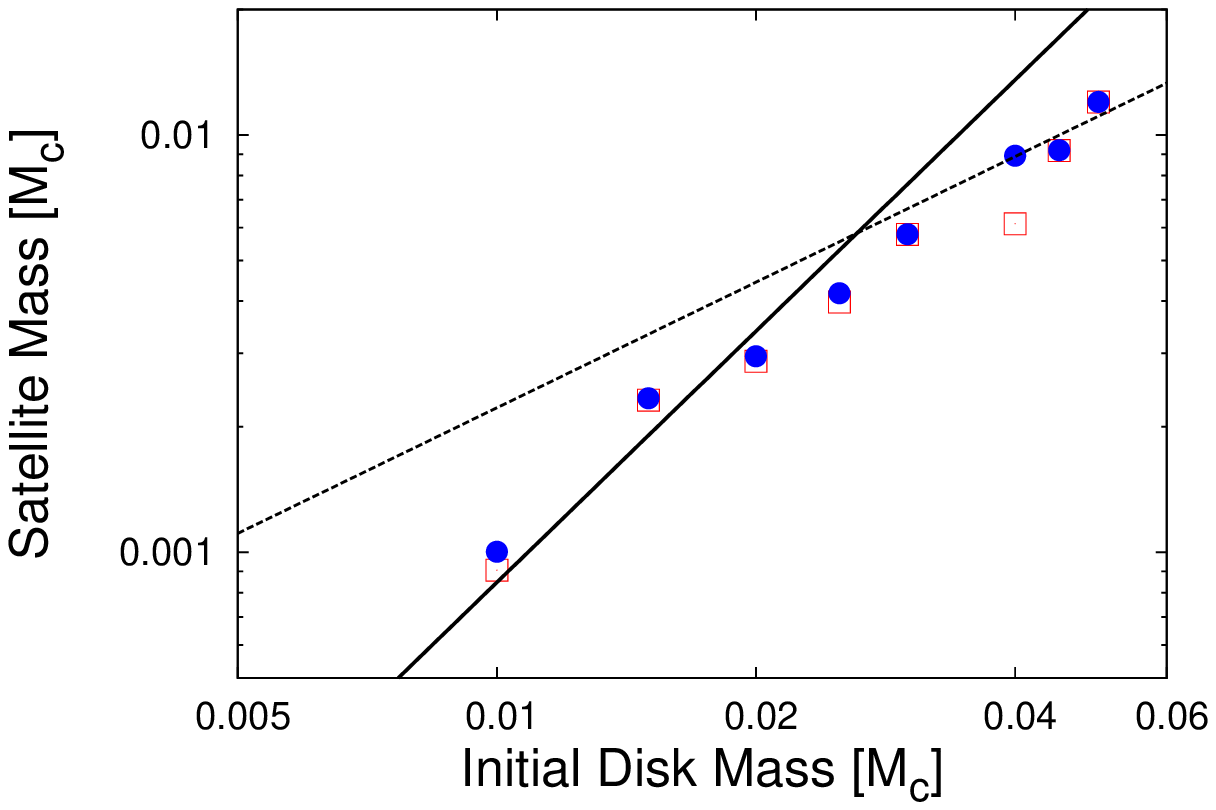}
\caption{Mass of the first satellite as a function of the initial mass of the disk ($j_{\rm disk,ini}=0.775$). Open 
		squares represent the mass of the first satellite, while the filled circles show 
		the sum of the masses of the first satellite and the second largest body in the 
		system (i.e., a co-orbital satellite or a smaller satellite on an outer orbit). The 
		dashed line represents the logarithmic fit to the numerical results for $M_{\rm disk,ini}/M_{\rm c} > 0.03$ 
		assuming that the satellite mass is proportional to the initial disk 
		mass. The solid line is the logarithmic fit to the results for $M_{\rm disk,ini}/M_{\rm c} < 0.03$,  
		assuming that the satellite mass is proportional to the square of the initial disk 
		mass. The masses of the satellites in the case of the formation of single-satellite 
		systems are those at the end of each simulation (i.e., those at $t = 500T_{\rm K}$), 
		while those in the case of the formation of multiple-satellite systems are those 
		just before the formation of the second satellite.}
\label{Ms_Md}
\end{figure}

\begin{figure}
\epsscale{.60}
\plotone{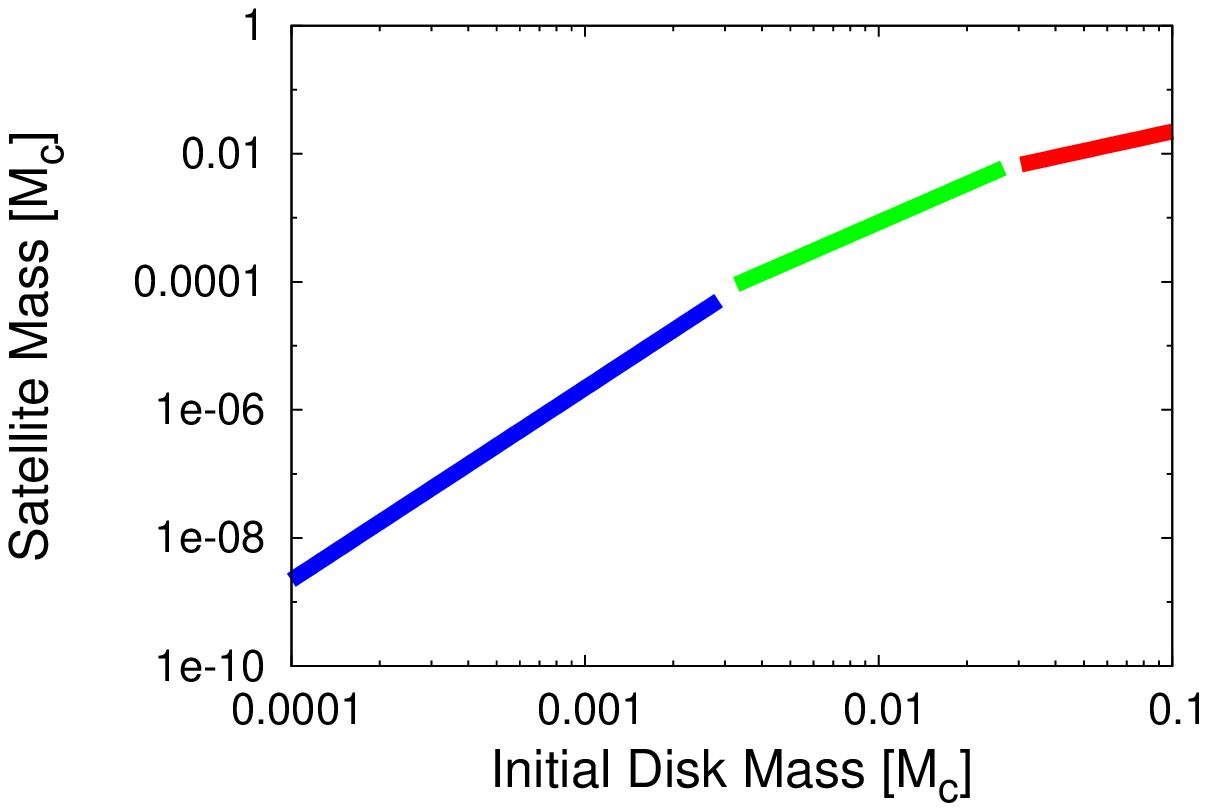}
\caption{Dependence of the mass of the first satellite on the initial mass of the disk 
		in three different regimes. The red line represents the case of the formation 
		of single-satellite systems from massive disks, where the satellite mass is 
		proportional to the disk mass. 
		The green line represents the case of the formation of two-satellite systems,   
		where the satellite mass is approximately proportional to the square of the 
		disk mass. These two lines were obtained by logarithmic fits to our numerical results (Figure \ref{Ms_Md}). 
		Typical results of the previous N-body simulation of lunar accretion from
		massive disks fall within the width of the red line 
		(i.e., $0.15 \lesssim M_{\rm s}/M_{\rm disk,ini} \lesssim 0.35$ for
		$0.7 \lesssim j_{\rm disk,ini} \lesssim 0.85$; \cite{Kokubo00}), while
		our numerical results for the formation of two-satellite systems
		fall within the width of the green line (Figure \ref{Ms_Md}).
		The blue line shows the result of \cite{CC12} on the mass of the satellite 
		formed at the end of “the discrete regime” (i.e., $M_{\rm s}/M_{\rm c} \simeq 2200(M_{\rm disk}/M_{\rm c})^3$). 
		Although they assume that the disk surface density is constant, actual disks should have surface density 
		distribution, and the mass of satellites accreted from such light disks should also 
		depend on the surface density (or angular momentum) distribution similarly to the numerical results 
		in the other two regimes. Note that actual transitions between the regimes are expected to be rather smooth.}  
\label{differents}
\end{figure}

\clearpage
\begin{figure}
\epsscale{.80}
\plotone{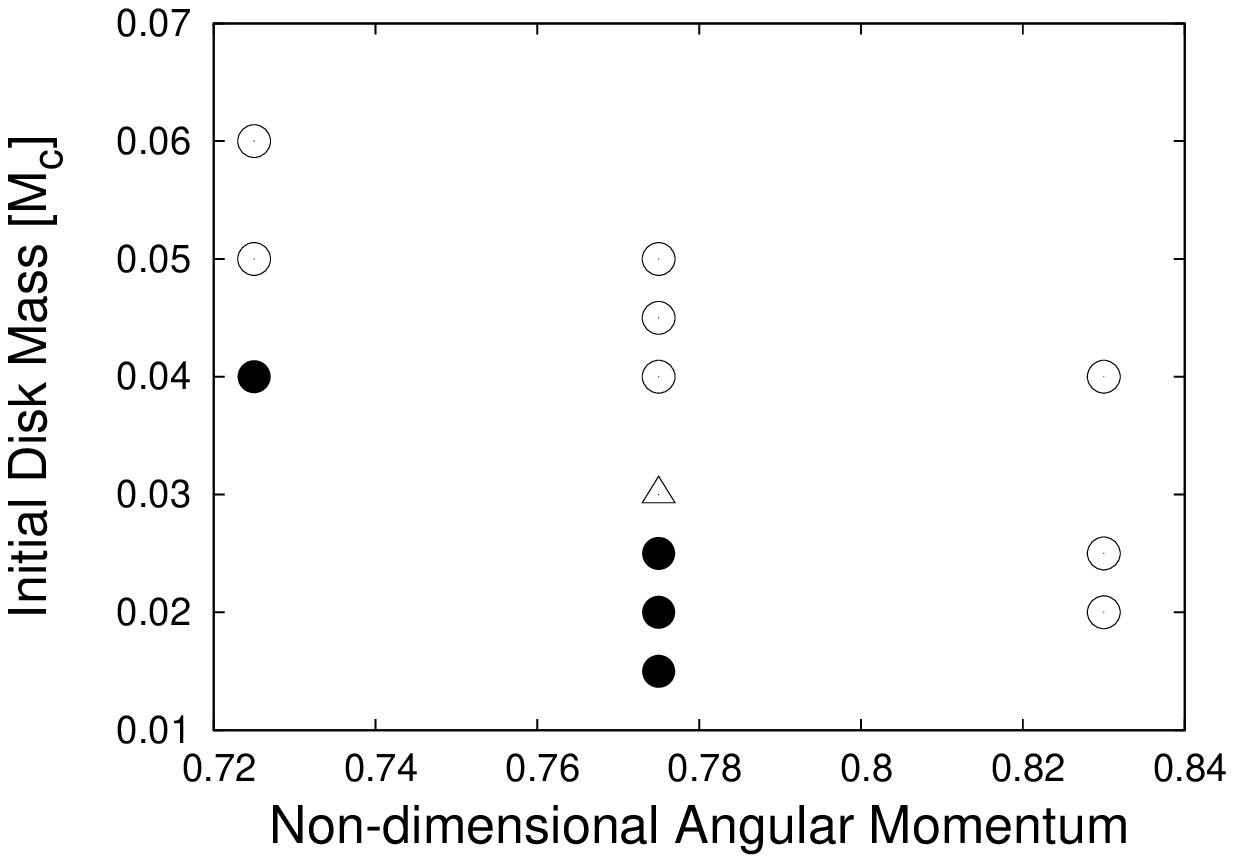}
\caption{Outcomes of simulations with various values of initial disk mass and non-dimensional 
		angular momentum of the disk. Filled circles indicate the case where the second satellite 
		is formed, while open circles represent the case that results in single-satellite 
		systems. Open triangle shows the marginal case where whether the second satellite is 
		formed or not depends on the choice of random numbers for generating initial positions and velocities of particles.}
\label{ms1_or_w_ms2}
\end{figure}

\clearpage
\begin{figure}
\epsscale{.80}
\plotone{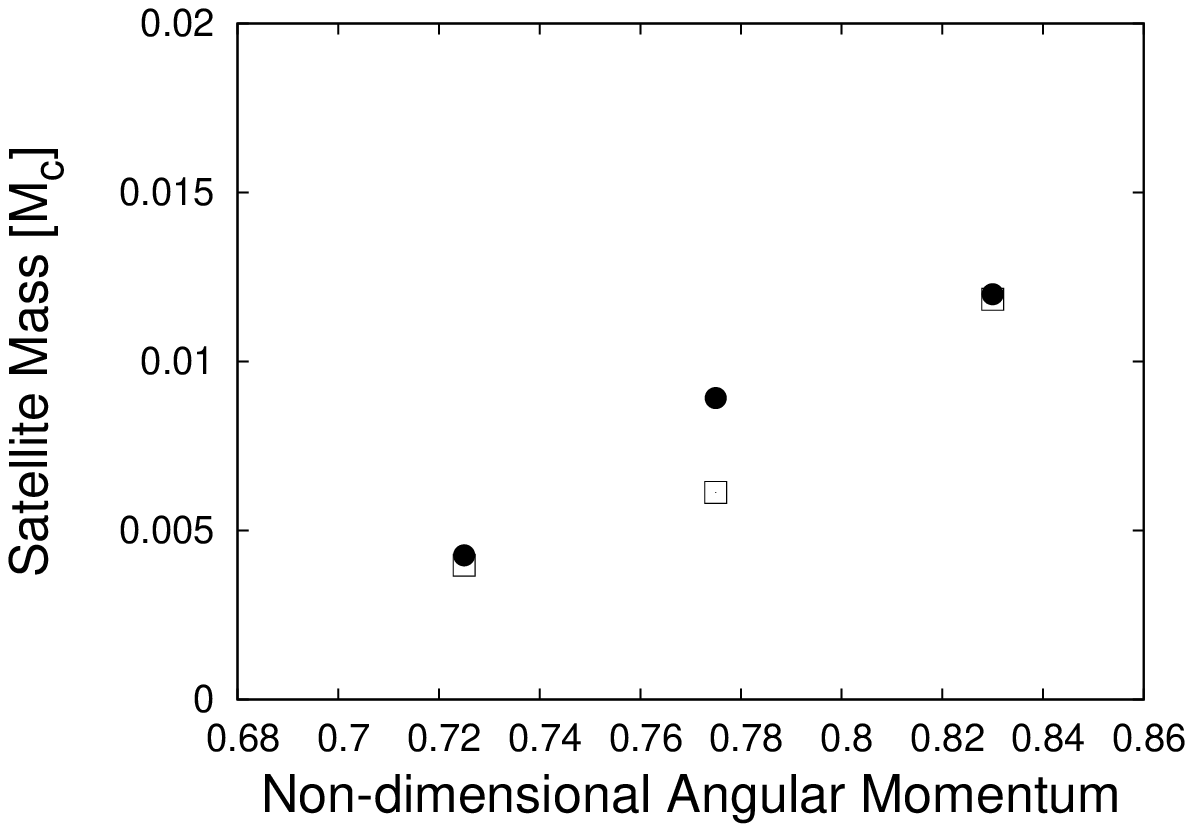}
\caption{Dependence of the mass of the first satellite on the non-dimensional angular 
		momentum of the initial disk (Runs 3, 6, and 12; $M_{\rm disk,ini}/M_{\rm c} = 0.04$ in all the 
		three cases). Open squares represent the mass of the first satellite, while the 
		filled circles show the sum of the masses of the first satellite and the second 
		largest body in the system (i.e., a co-orbital satellite or a smaller satellite on an outer orbit).}
\label{Ld_dependence}
\end{figure}

\clearpage
\begin{figure}
\epsscale{.70}
\plotone{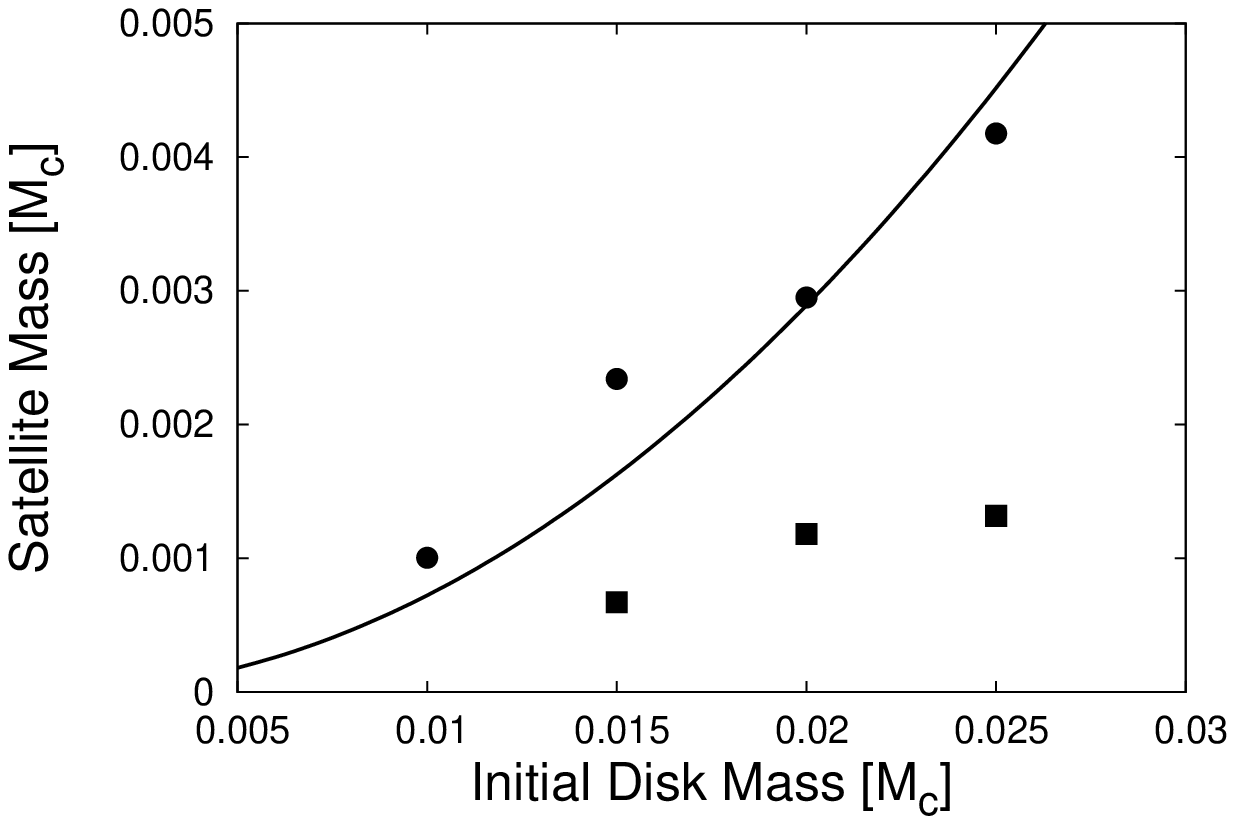}
\caption{Mass of the first satellite (circles) and the second satellite (squares) as a 
		function of the initial mass of the disk ($j_{\rm disk,ini}=0.775$). The solid line represents the fit to the 
		results for the first satellite, assuming that the satellite mass is proportional to 
		the square of the initial disk mass.}
\label{Ms1Ms2}
\end{figure}

\clearpage
\begin{figure}
\epsscale{.60}
\plotone{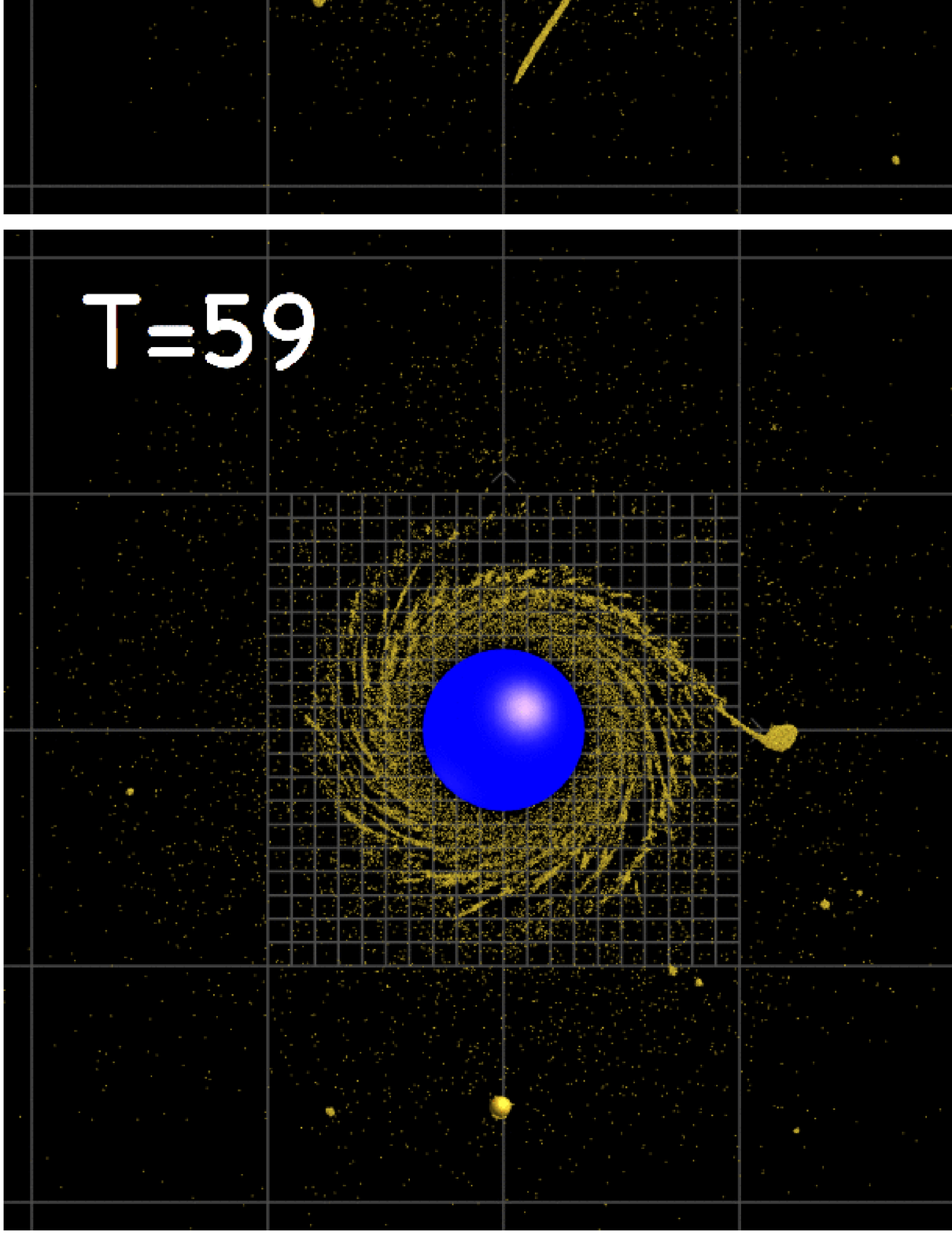}
\caption{Snapshots of the simulation for the case of the formation of the two-satellite 
		system with a larger satellite on an inner orbit and a smaller one on an outer 
		orbit (Run-7b; $M_{\rm disk,ini}/M_{\rm c} = 0.03$, $j_{\rm disk,ini} = 0.775$).}
\label{Md0.03_diversity}
\end{figure}

\clearpage
\begin{figure}
\epsscale{.45}
\plotone{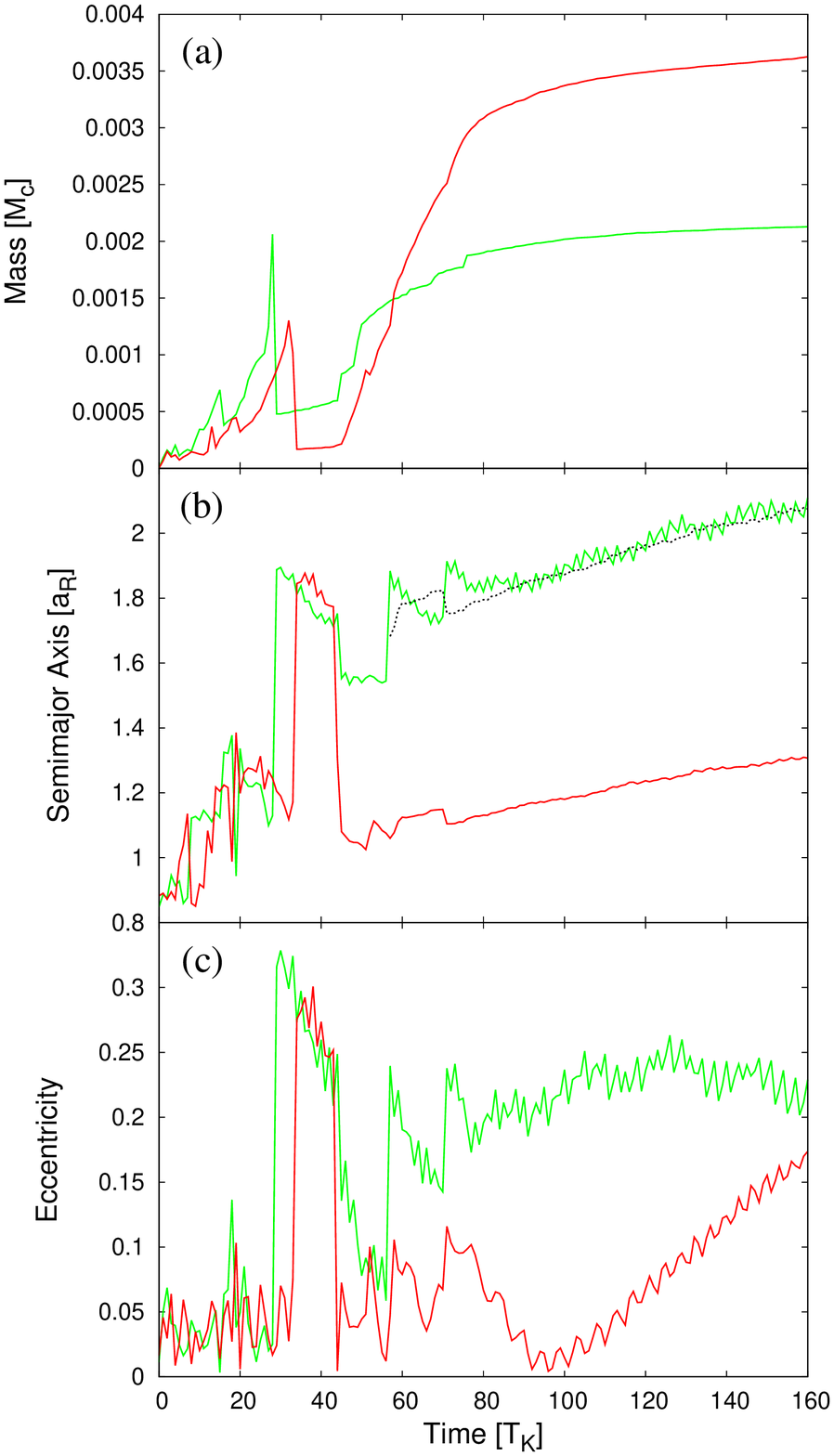}
\caption{Evolution of the mass (top panel), semi-major axis (middle panel), and 
		eccentricity (bottom panel) of the satellites in the case of Run-7b shown in 
		Figure \ref{Md0.03_diversity}. The red lines represent the one that eventually becomes the largest 
		satellite on an inner orbit, and the green lines represent the one that becomes 
		the smaller satellite on an outer orbit. The black dotted line in the middle panel 
		shows the radial location of the 1:2 mean motion resonance with the inner largest satellite.}
\label{diverse}
\end{figure}

\clearpage
\begin{table}
\begin{center}
\caption{Initial Disk Parameters and Mass of Formed Satellites\vspace{0.5em}\label{tbl-1}} 
\begin{tabular}{crrrrr}
\tableline\tableline
Run & $M_{\rm disk,ini}/M_{\rm c}$ & $j_{\rm disk,ini}$ & $N$ & $M_{\rm s,out}/M_{\rm c}$ & $M_{\rm s,in}/M_{\rm c}$ \\
\tableline
1  &0.060  &0.725 &30,000 &8.72$\times10^{-3}$ &$\ldots$\\
2  &0.050  &0.725 &30,000 &7.98$\times10^{-3}$ &$\ldots$\\
3  &0.040  &0.725 &50,000 &$^*$4.27$\times10^{-3}$ &2.43$\times10^{-3}$\\
4  &0.050  &0.775 &30,000 &1.20$\times10^{-2}$&$\ldots$\\
5  &0.045  &0.775 &30,000 &9.18$\times10^{-3}$&$\ldots$\\
6  &0.040  &0.775 &30,000 &$^*$8.92$\times10^{-3}$&$\ldots$\\
7  &0.030  &0.775 &50,000 &5.78$\times10^{-3}$ &$\ldots$\\
7b &0.030  &0.775 &50,000 &2.13$\times10^{-3}$ &3.63$\times10^{-3}$\\
8  &0.025  &0.775 &50,000 &$^*$4.18$\times10^{-3}$&1.32$\times10^{-3}$\\
9  &0.020  &0.775 &50,000 &2.87$\times10^{-3}$&1.18$\times10^{-3}$\\
10 &0.015  &0.775 &50,000 &2.32$\times10^{-3}$&6.71$\times10^{-4}$\\
11 &0.010  &0.775 &50,000 &$^*$1.00$\times10^{-3}$&$\ldots$\\
12 &0.040  &0.830 &30,000 & 1.18$\times10^{-2}$&$\ldots$\\
13 &0.025  &0.830 &30,000 & 7.81$\times10^{-3}$&$\ldots$\\
14 &0.020  &0.830 &30,000 & 5.01$\times10^{-3}$&$\ldots$\\
\tableline
\end{tabular}
\end{center}
{\it Note}: In the case where only one satellite is formed, $M_{\rm s,out}$ represents its 
mass at the end of simulation; when a co-orbital satellite and/or a small moonlet 
on an outer orbit is also formed as the second largest body in the system, their 
masses are also included in $M_{\rm s,out}$ to facilitate comparison with other cases. In 
the case where two satellites are formed, $M_{\rm s,in}$ and $M_{\rm s,out}$ are the masses of 
satellites on the inner orbit and the outer orbit, respectively; in this case, $M_{\rm s,in}$  
is obtained from the mass of the second satellite just after its rapid growth 
phase. The asterisks denote the cases where a co-orbital satellite was formed 
on the orbit of the primary satellite.
\end{table}

\end{document}